\RequirePackage{fix-cm}
\documentclass[]{svjour3}                     % onecolumn (standard format)
\smartqed  % flush right qed marks, e.g. at end of proof
\usepackage{pdflscape}
\usepackage{fixltx2e}
\usepackage{xcolor}
\usepackage{hyperref}
\usepackage{graphicx}
\usepackage{amsmath}
\usepackage{multirow}
\usepackage{tabularx}
\usepackage{amsfonts}
\usepackage{amssymb}
\usepackage{multirow}
\usepackage{epstopdf}
\usepackage[caption=false]{subfig}
\usepackage{float}
\usepackage[utf8]{inputenc}
\usepackage{caption}
\usepackage{algorithm}
\usepackage{ amssymb }
\usepackage{algpseudocode}
\usepackage[numbers]{natbib}
\usepackage{tikz}
\newcommand\mybox[2][]{\tikz[overlay]\node[fill=yellow!20,inner sep=2pt, anchor=text, rectangle, #1] {#2};\phantom{#2}}

\journalname{}
\begin{document}
	
	\title{CoSec-RPL: Detection of Copycat Attacks in RPL based 6LoWPANs using Outlier Analysis}
	
	\titlerunning{Detection of non-spoofed copycat attacks}        % if too long for running head
	
	\author{Abhishek Verma$^{1,*}$\and
		Virender Ranga$^{1}$
	}

	\institute{$^{1}$Department of Computer Engineering, National Institute of Technology Kurukshetra, India \\
		\email{$^{*}$abhiverma866@gmail.com} \\ 
		\email{virender.ranga@nitkkr.ac.in} \\         %  \\
	}
	
	\date{Received: xx-xx-xxxx / Accepted: xx-xx-xxxx}
	\maketitle
\begin{abstract}
	
	The IPv6 Routing Protocol for Low-Power and Lossy Networks (RPL) is the standard routing protocol for IPv6 based Low-Power Wireless Personal Area Networks (6LoWPANs). In RPL protocol, DODAG Information Object (\textcolor{black}{DIO}) messages are used to disseminate routing information to other nodes in the network. A malicious node may eavesdrop DIO messages of its neighbor nodes and later replay the captured DIO many times with fixed intervals. \textcolor{black}{In this paper, we present and investigate one of the severe attacks named as a non-spoofed copycat attack, a type of replay based DoS attack against RPL protocol. It is shown that the non-spoofed copycat attack increases the Average End-to-End Delay (AE2ED) and Packet Delivery Ratio (PDR) of the network.} 
	%Copycat attack may severely affect the Quality of Service (QoS) of IoT applications. 
	Thus, to address this problem, an Intrusion Detection System (IDS) named CoSec-RPL is proposed in this paper. \textcolor{black}{The attack detection logic of CoSec-RPL is primarily based on the idea of Outlier Detection (OD)}. 
	%The effectiveness of CoSec-RPL is analyzed by performing an in-depth experimental study.
	\textcolor{black}{CoSec-RPL significantly mitigates the effects of the non-spoofed copycat attack on the network's performance.
		The effectiveness of the proposed IDS is compared with the standard RPL protocol. }
	\textcolor{black}{The experimental results indicate that CoSec-RPL detects and mitigates non-spoofed copycat attack efficiently in both static and mobile network scenarios without adding any significant overhead to the nodes.} To the best of our knowledge, CoSec-RPL is the first RPL specific IDS that utilizes OD for intrusion detection in 6LoWPANs.\footnote{The final publication is
		available at https://link.springer.com/article/10.1007/s11235-020-00674-w}
\end{abstract}


\section{Introduction}\label{Introduction}
In the recent years, Internet of Things (IoT) has been a major player among various evolving networking paradigms  \cite{ashton2009internet, IoTCore1, ammar2018internet}. \textcolor{black}{International Data
	Corporation (IDC) predicted that there will be 41.6 billion connected IoT devices worldwide by 2025} \cite{idc2}. While, worldwide spending on IoT is expected to cross the \$1 trillion mark in 2022 \cite{IDC}. 
With this much expansion of IoT, the security issues related to it are also expanding. The increase in the number of IoT devices also increases the number of incredible risks. These risks primarily include users' security and privacy getting exposed to cyber attacks \cite{Raoof, Alaba2017, AIREHROUR2016198, ziegeldorf2014privacy, Yang7902207}. In the present scenario, many IoT applications are deployed on IPv6 over Low-Power Wireless Personal Area Networks (6LoWPANs). The 6LoPWAN concept enables Internet  Protocol (IP) on tiny devices, i.e., embedded devices with limited processing power, small onboard memory, and limited energy resources. 6LoWPAN is based on Low Power and Lossy Networks (LLNs), which have high packet loss and low throughput communication links \cite{COLAKOVIC2018, winter2012rpl}. LLNs are realized by resource constrained devices which operate on very low power, to support longer network lifetime  \cite{Musaddiq}. The characteristics of LLNs like resource constrained nature, high packet loss, and low network throughput make traditional routing protocols unsuitable for LLN \cite{tripathi2014design}. To solve this issue, the IPv6 Routing Protocol for Low-Power and Lossy Networks (RPL) was standardized (RFC $ 6550 $) \cite{winter2012rpl}. The RPL protocol provides energy efficient routing in LLNs. However, the RPL protocol remains exposed to various cyber attacks, which may jeopardize users' security and privacy \cite{RIAHISFAR2018118, Verma2019, abhishek_verma_IoTSIU, verma2019addressing}. \textcolor{black}{The critical applications like healthcare and smart grid, when becoming the target of such threats, may result in life-threatening incidents. This motivated us to explore and perform an in-depth analysis of one such threat (i.e., copycat attack) and design a defense mechanism to detect and mitigate it.} The vulnerabilities and threats associated with the RPL protocol have been rigorously studied by cyber security researchers \cite{VermaIEEE, adat2018security}. \textcolor{black}{In this paper, the main focus is on a replay mechanism based routing attack which is known as copycat attack, that affects the Quality of Service (QoS) of real-time wireless networks.}

According to standard RPL specification (RFC 6550), the RPL protocol supports a secure mode to provide integrity and confidentiality to data and control packets. The secure mode incorporates traditional cryptography mechanisms to enable security and privacy \cite{ghaleb2018survey}. However, the standard RPL specification does not specify any details of secure key management, which restricts the usage of cryptography in resource constrained devices \cite{malik2019survey, seeber2013towards, Perazzo2017a}. Moreover, the traditional cryptography based security methods (e.g., Public-key and     Symmetric-key cryptography) consume a lot of computing resources, degrade the network's performance, and reduce the lifetime of IoT networks \cite{shamsoshoara2019survey}. The RPL protocol is unprotected from cyber attacks (e.g., routing attacks), where an attacker node can exploit its vulnerabilities to compromise the legitimate nodes. The attacks may degrade the network's overall performance significantly, which consequently, affects the operation of IoT applications. \textcolor{black}{One such destructive attack is termed a copycat attack, a type of replay mechanism based direct attack that targets the legitimate node's resources.} It is a Denial-of-Service (DoS) attack, which has the ability to \textcolor{black}{severely} degrade the performance of 6LoWPANs. To launch this attack, an attacker node eavesdrops DODAG Information Object (DIO) messages of legitimates neighbor nodes, and later sends the previously eavesdropped DIO messages many times with fixed replay interval. In this manner, the attacker introduces a high level of congestion and interference in the network, which leads to the creation of sub-optimized routes. Moreover, the attack also forces nodes to transmit DIO messages unnecessarily and performs unessential routing related operations. The copycat attack can be achieved even without stealing cryptography keys of legitimate nodes, which gives a significant advantage to the attacker (outsider attack scenario). Moreover, an attacker does not need to have any high range radio antenna or any other specialized hardware to perform copycat attacks.   

The major problem with 6LoWPANs is that the resource constrained devices lack built-in security. Moreover, the RPL protocol does not have any inbuilt Intrusion Detection System (IDS) to provide any defense against cyber attacks. Most importantly, there is no built-in security mechanism to provide defense against routing attacks, which are very common in wireless networks. Therefore, in this research paper, a new Outlier Detection (OD) \cite{kumar2016anomaly, jabez2015intrusion} based IDS named as CoSec-RPL (abbreviation of ``copycat secured RPL protocol") is proposed to detect copycat attack. CoSec-RPL detects the malicious neighbors and blocks all further communications from it. The main idea behind our proposed IDS is to use the OD mechanism to detect neighbors with abnormal behavior. CoSec-RPL has five major advantages. \textcolor{black}{Firstly}, it does not introduce any communication overhead. \textcolor{black}{Secondly}, it does not require any good network trace for model training. \textcolor{black}{Thirdly}, its performance improves with time. \textcolor{black}{Fourthly}, it does not impose any significant memory overhead on the nodes. \textcolor{black}{Fifthly}, it can be easily extended to detect other RPL specific routing attacks. Major contributions of the paper can be summarized as:

\begin{enumerate}
	\item \textcolor{black}{The impact of copycat attacks on RPL is analyzed through simulations.}
	\item \textcolor{black}{An IDS, named CoSec-RPL, targeting non-spoofed copycat attacks is presented and verified through simulations. }  
\end{enumerate}

The next section of this paper presents a brief overview of RPL protocol, copycat attack, and outlier detection. In Section \ref{Section:Related Work}, relevant works are discussed. The proposed solution is described in Section \ref{Section:ProposedSolution}. A discussion on performance evaluation of the proposed solution is presented in Section \ref{Section:PerformanceEval}. Some possible extensions of proposed solution are discussed in Section \ref{Section:CoSec-RPL extensions}, and finally we conclude the paper in Section \ref{Section:Conclusion}.

\section{Background}\label{Section:Background}
In this Section, we describe the RPL
protocol, copycat attack, and outlier detection. 
\subsection{Overview of RPL Protocol}\label{Section:Overview of RPL}

\textcolor{black}{In this section, building elements, control messages, and fault tolerance mechanisms of the RPL protocol are discussed. } 

\subsubsection{Building Elements of RPL}
\begin{itemize}
	\item DODAG: RPL is founded on the idea of Directed Acyclic Graphs(DAGs) \cite{RPLNutshell}. In RPL, the IoT devices are logically interconnected with each other using mesh and tree topology. In Destination Oriented Directed Acyclic Graph (DODAG), the root node (gateway) acts as an interface between 6LoWPAN nodes and the Internet. A network may contain more than one DODAG, which collectively forms an RPL Instance and uniquely identified by RPL Instance ID. In a network, more than one RPL Instance may run at a time. An RPL node may be associated with only one DODAG per RPL Instance. Each node of a DODAG is assigned a rank which represents ``the node's individual position relative to other nodes with respect to a DODAG root" \cite{winter2012rpl}.  The rank concept is implemented in RPL: (1) to detect and avoid routing loops; (2) to build parent-child relationships; (3) to provide a mechanism for nodes to differentiate between parent and siblings; (4) to enable nodes to store a list of preferred parents and siblings which can be utilized during link repair. DODAG is built during the network topology setup phase, during which each node uses RPL control messages to find the optimal set of parents towards the root and link itself with the preferred parent. The selection of preferred parents is based on an Objective Function (OF). The OF defines the procedure for rank computation from routing metrics and selection of optimal routes in DODAG. RPL may use different OF as per the application's requirement. Some common OF are ETX Objective function \cite{ETXOF}, Minimum Rank with Hysteresis Objective Function (MRHOF) \cite{MRHOF}, and Objective Function Zero (OF0) \cite{OF0}. RPL supports Multi-Point-to-Point, Point-to-Multipoint, and Point-to-Point \cite{RPLNutshell,medjek2018security} network topologies. \textcolor{black}{An example of RPL DODAG with N Nodes having IPv6 addresses range from aaaa::1 to aaaa::N is shown in Fig. \ref{DODAG}.}
	
	\begin{figure}
		\centering
		\includegraphics[width=.6\textwidth]{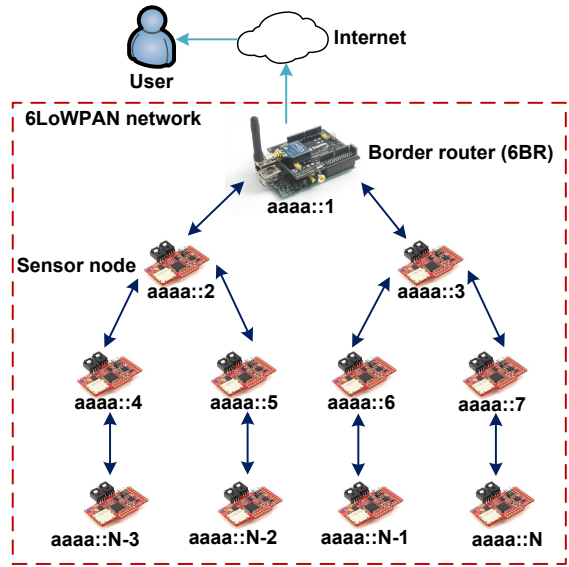}
		\caption{An example of RPL DODAG with N nodes}
		\label{DODAG}
	\end{figure}
	
	\item Control Messages: RPL defines a new category of ICMPv6 control messages known under Type 155 and defined in \cite{winter2012rpl, 155IANA}. RPL control messages include DODAG Information Object (DIO), DODAG Information Solicitation (DIS), Destination Advertisement Object (DAO), and Destination Advertisement Object Acknowledgment (DAO-ACK).
	DIO message carries routing information relevant to existing DODAG and allows other nodes to find an RPL instance and its configuration parameters. Also, it enables a node to select its preferred parent set and performs DODAG maintenance. DIS message is used to solicit a DIO message from an RPL node. It is used by the new or existing node to search for a nearby DODAG. DAO message is used to forward downward route information in the upward direction along the DODAG, finally reaching the root node. DAO-ACK message is a unicast packet send an acknowledgment by a DAO parent or DODAG root, in reply to a unicast DAO message \cite{RPLNutshell}. 
	
	\item Trickle Timer: RPL uses an adaptive timer mechanism called as ``Trickle timer" in order to limit control traffic in the network \cite{levis2011trickle}. RPL uses a dynamic mechanism to control the number of DIO messages sent by the resource-constrained nodes for minimizing energy consumption. Trickle timer decides when a node should multicast the DIO messages, and it gets reset in case of inconsistency detection in the network, i.e., loops and link loss, change in parent set, etc. The interval of the trickle timer is increased, decreased in case of a stable network and inconsistency detection, respectively. In the case of a stable topology, the trickle timer interval is increased. Thus, the number of DIO sent are decreased, and when this interval is decreased, the number of DIO sent are increased in order to fix the inconsistency issue \cite{IPSOAllianceRPL}.
	
\end{itemize}
\subsubsection{Fault Tolerance Mechanisms}
RPL defines some important network management mechanisms. \textcolor{black}{It fulfills self-healing characteristics by incorporating a DODAG repair mechanism (global and local), which are triggered during inconsistency detection, loop detection, and avoidance mechanisms to handle routing loops.} Inconsistencies include node failure, link failures, change in parent set, and routing loops. A loop may occur when a node, after losing all its parent, joins another node (makes parent) that was earlier in its sub-DODAG. Loop avoidance and loop detection mechanisms of RPL are contrary to those applied in traditional IP networks \cite{xie2010routing}. In this section, various fault tolerance mechanisms are discussed. 

\begin{itemize}
	\item Loop Avoidance: RPL defines two strict rules based on a rank property for avoiding loops in the network. The first rule is termed as ``max\_depth rule". It states that a node must not select a neighboring node as its parent whose rank is higher than its own rank. The second rule states that a node must not increase its rank by selecting nodes of higher rank as their preferred parent in order to increase its parent set size. 
	
	\item Loop Detection: Since loops are unavoidable in LLNs, hence the need for loop detection mechanisms arises. RPL defines a mechanism to detect routing loops whenever they occur. A data path validation mechanism is used by RPL to resolve routing loops. It involves setting and processing some specific bits contained in the RPL routing header. RPL ensures that the packets moving in the wrong direction are detected as a part of some loop. Loop recovery mechanism further involves resetting of trickle timer for repairing network topology while discarding packets being received at that time. 
	
	\item Local Repair: \textcolor{black}{RPL triggers the DODAG local repair mechanism in case of a node failure, link failure, and loop detection. Local repair aims to rapidly find an alternate parent/path (may not be optimal) without putting any global implication on entire DODAG.}  
	
	\item Global Repair: \textcolor{black}{When local repairs are found to be inefficient while performing network recovery as they start diverging DODAG to a non-optimal state due to the presence of many inconsistencies, then the whole DODAG needs to be rebuilt from scratch}. Global repair is performed by incrementing DODAG version number, which leads to the reconstruction of the whole DODAG, where nodes recompute their rank to form an optimal topology \cite{hui2012routing}.     
\end{itemize}

\subsubsection{RPL Modes}
Two modes of operations are supported by the RPL protocol in order to maintain downward routes. \textcolor{black}{In this section, storing and non-storing mode of the RPL protocol \cite{winter2012rpl, RPLNutshell} are highlighted.}
\begin{itemize}
	\item Storing mode: \textcolor{black}{In the storing mode, downward routes start to propagate from leaf nodes to root node through intermediate router nodes.} Every child node sends DAO message to its parent who initially stores information contained in that message and later sends a new DAO message containing aggregated reachability information to its parent. Thus, each node knows the path to every other node in the RPL network.
	
	\item Non-storing mode: In the non-storing mode, leaf nodes unicast DAO message to the DODAG root node. Unlike storing mode, intermediate router nodes do not store any information from DAO message; instead, they only append their address to it and forward to the parent. It is done to form a reverse routing path. Thus, only the DODAG root knows a path to every node in the network.    
\end{itemize}  

\subsection{Copycat Attack}\label{Section:Copycat Attack}

The main target of the copycat attack is to degrade the routing performance of RPL based 6LoWPANs so that the QoS of real-time applications gets affected. In this, an attacker may compromise a legitimate internal node and reprogram it to introduce the high level of congestion and interference in the network. The attacker can also choose an outsider attack strategy to perform this attack. To launch a copycat attack, an attacker eavesdrops the DIO messages of nearby nodes, and later sends (multicast) the captured DIO message (with or without modification) many times with a fixed replay interval. \textcolor{black}{The copycat attack can be of two types: 1) non-spoofed; 2) spoofed.} In \textcolor{black}{``non-spoofed copycat attack"}, the eavesdropped DIO is sent after modifying the source IP of the ICMPv6 packet containing DIO message. The attacker sends the unmodified captured DIO with its own IP address in the ICMPv6 packet, which forces the receiving (victim neighbors) nodes to believe that the packet is from a legitimate sender and makes them perform unnecessary routing related operations. Therefore, an attacker is able to drain victim's resources and disrupts its normal packet forwarding behavior. The second type of copycat attack is termed as \textcolor{black}{``spoofed copycat attack"}. In this attack, the eavesdropped DIO is sent to neighbor nodes after replacing the source IP address of encapsulating IPv6 packet with the IP address of legitimate DIO sender, i.e., the sender of eavesdropped DIO message. This makes the receiver believe that the sender of DIO is its in-range neighbor. The victim nodes may even try to add the out of range neighbor, assuming that it leads to the optimal route to the gateway. \textcolor{black}{In simple words, in non-spoofed copycat attack the adversary uses its IP address as the source, and in spoofed copycat attack the adversary uses the source IP address of a legitimate node as a source.} Both types of attacks introduce heavy congestion and interference in their attack region, which consequently, decreases the Packet Delivery Ratio (PDR) and increases the Average End-to-End Delay (AE2ED) of the underlying network. The main difference between copycat attack and other replay attack variants (i.e., routing information replay and neighbor attack) lies in the frequency of replaying the packets and the packet field being modified. In other RPL specific replay attacks, the attacker primarily aims to introduce the un-optimized or non-existing paths in the network by merely replaying the previously eavesdropped DIO packet after a certain period of time. In contrast, the copycat attacker focuses on the combination of the replay and interference method. The attack also forces legitimate nodes to make unnecessary DIO transmissions, which consequently, increases the control packet overhead of the network.

\begin{figure}[!h]
	\centering
	\includegraphics[width=.8\textwidth]{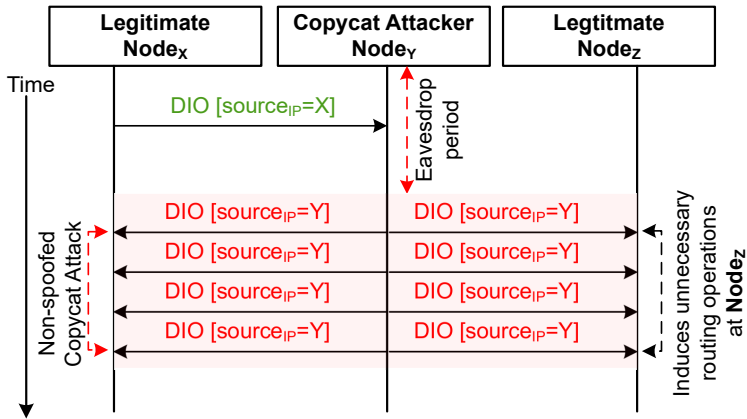}
	\caption{\textcolor{black}{Attacker's approach for launching non-spoofed copycat attack}}
	\label{Fig:CopycatFlow}
\end{figure}

Moreover, the standard RPL specification states that the link quality (e.g., Expected Transmission Count) must be computed before adding a new node in the candidate parent set when MRHOF is used. Upon receiving the replayed DODAG Information Object (DIO) messages, a probing mechanism is initiated to asses the link quality. In this case, the probing fails because the replayed source is not in the communication range of the node, hence the path is assumed to be bad and consequently, discarded \cite{wallgren2013routing}. Thus, the neighbor attack is ineffective if the nodes use ETXOF or MRHOF. Moreover, when an eavesdropped packet is frequently replayed multiple times with a fixed interval, a heavy interference is introduced in the network region, i.e., an attacker's communication range. \textcolor{black}{Also, copycat attack with fixed time interval keeps node busy continuously and consequently, degrades the network's performance. It is to be noted that a copycat attack can also be performed with random intervals. However, the interval needs to be short in order to perform maximum damage to the network. Also, adding a mechanism to compute random interval very frequently will impose computational overhead to the attacker node, thereby decreasing attacker node’s lifetime. Considering the fact that the attacker's primary target is to cause maximum damage to the network, it will simply choose a shorter interval (fixed value) and perform attack for longer time.  In this study, we have considered the attack with fixed intervals. Analysis and detection of copycat attack with random intervals will be considered in our future work.” } The copycat attacker node is programmed in such a way that it remains isolated (neither makes a parent nor becomes a parent) from the network while only performing a replay attack. In this way, an attacker is able to reduce its own energy consumption rate for performing a long-lasting attack. \textcolor{black}{In case of spoofed copycat attack where an attacker uses source IP of one or more legitimate nodes (i.e., like Sybil attack), the attack will be ineffective if RPL is configured with MRHOF. Whereas in case the RPL is configured with OF0, then the attacker will succeed in persuading legitimate nodes that it is a potential parent. This is because the nodes do not check for neighbor reach-ability in case of OF0.} In this paper, we have focused on \textcolor{black}{non-spoofed} copycat attack, and proposed an IDS to detect such attacks in RPL based 6LoWPANs. The \textcolor{black}{non-spoofed} copycat attack is illustrated in Fig. \ref{Fig:CopycatFlow}.

%\section{System Model and Assumptions}\label{systemModel}

\subsection{Outlier Detection}\label{Section:Outlier detection}
An outlier is defined as ``an observation (or subset of observations) which appears to be inconsistent with the remainder of that set of data" \cite{barnett1974outliers}. OD involves the detection and removal of outliers from the data. OD has been \textcolor{black}{commonly} used for a long time to detect anomalies present in the data. Outliers can arise in data due to intentional or unintentional software and hardware errors, e.g., data entry error. In machine learning, removing outliers is one of the primary tasks in data preprocessing to leverage the quality of a prediction or classification model. Indeed, OD is important to any quantitative discipline that needs a good quality of data. There are many OD methods available in the literature, and the most popular one is known as the Interquartile Range (IQR). The standard deviation around the mean can be used to detect outliers. However, mean and standard deviation are sensitive to outliers and may lead to incorrect results. This problem is solved by the IQR method as it uses the median instead of the mean.

The IQR is a measure of statistical dispersion based on dividing data into quartiles. The value of IQR represents the middle 50\% of sorted data (ascending). IQR is calculated as in Eq. \ref{Eq.IQR}, where $ Q3, Q1 $ are third and the first quartile, respectively.
$  $

\begin{equation}
IQR = Q3-Q1
\label{Eq.IQR}
\end{equation}

\begin{figure}[!h]
    	\centering
    	\includegraphics[width = .8\textwidth]{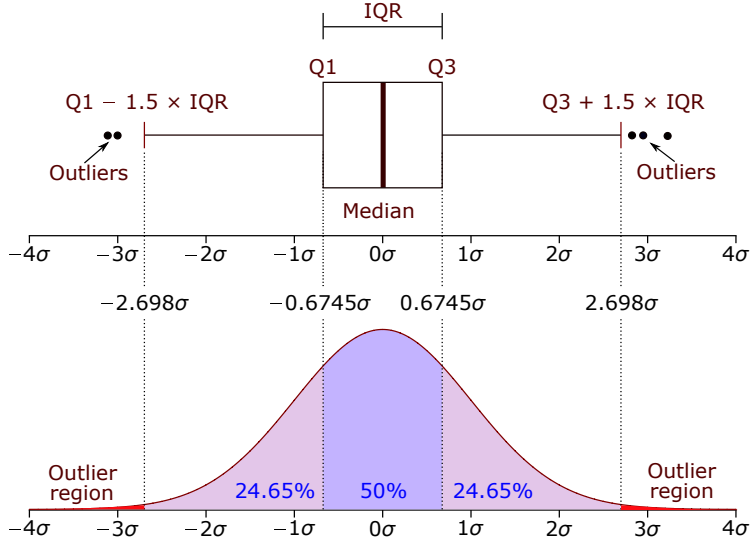}
    	\caption{Box plot and probability density function of a normal distribution}
    	\label{IQR}
\end{figure}

To determine the IQR,  firstly, the median ($ \tilde{x} $) of the data is computed. Then, the first quartile ($ Q1 $) and third quartile ($ Q3 $) are computed. Q1, Q3 are the median of the lower and upper half of the data. After the computation of $ Q1 $ and $ Q3 $, the IQR is computed using Eq. \ref{Eq.IQR}. In order to visualize the distribution of data for better analysis, box plots are used. Fig. \ref{IQR} illustrates an example of a box plot and probability density function of a normal distribution. The illustration visualizes the minimum, $ Q1 $, $ \tilde{x} $, $ Q3 $, and maximum value. Tukey \textit{et al.} proposed to use 1.5$ \times $IQR (Tukey fences) as a demarkation line for outliers \cite{hoaglin2003john}. As per 1.5$ \times $IQR rule, points below  \textit{Lower limit}, and points above \textit{Upper limit} are considered as outliers. The \textit{Lower limit}, \textit{Upper limit} are calculated as in Eqs. \ref{LL} and \ref{UL}, respectively. The OD problem can be mapped to the intrusion detection problem of RPL based 6LoPWANs. Where, an outlier can be a node with abnormal behavior (i.e., malicious node) which needs to be identified and eliminated for achieving better network performance. 

\begin{equation}
Lower~limit = Q1 - 1.5 \times IQR
\label{LL}
\end{equation}   

\begin{equation}
Upper~limit = Q3 + 1.5 \times IQR
\label{UL}
\end{equation}

%N(0, 1$ \sigma^2 $)        

\section{Related Work}\label{Section:Related Work}

Replay and flooding attacks have been widely studied by the WSN researchers and many IDS have been suggested for detection of such attacks \cite{mohammadi2019defending,hamid2006routing,raymond2008denial,pathan2006security,gungor2009industrial}. However, such solutions cannot be directly applied in RPL security because RPL protocol has different operating mechanisms and different format of control messages. In the literature, there are a limited number of works on the detection of replay and flooding attacks against RPL protocol.  Le \textit{et al.} \cite{Le2016} proposed a specification based IDS for detecting Rank \cite{le2013impact}, Local repair, Neighbor, DIS and Sinkhole attacks. The proposed IDS is based on the Extended Finite State Machine-generated from a semi-auto profiling technique. Tsao \textit{et al.} \cite{tsao2015security} suggested using a counter to ensure the freshness of the data and control packets for defending a replay attack. However, no experimental study on the behavior of suggested attacks, and no performance evaluation of the suggested solutions is done in this study. Verma \textit{et al.} \cite{verma2019addressing, verma2020mitigation} proposed a lightweight defense mechanism to secure RPL  against the DIS flooding attack. The authors used standard RPL parameters like DIS start delay and DIS interval to detect and mitigate the attack. There are several works that have addressed other routing attacks, particularly to RPL. Ghaleb \textit{et al.} \cite{ghaleb2018addressing} proposed a security mechanism which is known as SecRPL to secure RPL against the DAO falsification attack. An enhanced version of the RPL protocol is proposed by Ariehrour \textit{et al.} \cite{Airehrour2018} to detect rank and sybil attacks. Raza \textit{et al.} \cite{RAZA20132661} developed SVELTE for the detection of sinkhole, selective forwarding and spoofing attacks. Mayzaud \textit{et al.}\cite{Mayzaud2014} analyzed version number attacks and suggested a distributed monitoring scheme to detect such attacks \cite{Mayzaud2016, Mayzaud2016Version, Mayzaud2017}. Gara \textit{et al.} \cite{GaraInderscience} addressed Selective forwarding and clone ID attacks using a hybrid IDS based on the Sequential Probability Ratio Test with an Adaptive Threshold. Bostani \textit{et al.} \cite{Bostani2017} proposed a hybrid IDS based on the combination of anomaly and specification detection engines to detect sinkhole and selective forwarding attacks. The existing security solutions are not suitable for the detection of copycat attack because of different attack characteristics, i.e., the copycat attack is a combination of flooding and replay attack. Where, flooding attack induces heavy congestion and interference, while the replay attack sub-optimizes the DODAG. To the best of our knowledge, there are no RPL specific IDS present in the literature that is capable of detecting such an attack. 
 
 \begin{table}[!h]
 	\centering
 	\caption{An example of outlier detection using modified IQR method}
 	\label{tab:IQR_example}
 	\begin{tabular}{|c|c|c|c|c|c|c|c|c|c|c|c|c|}
 		\hline
 		\textbf{} & \multicolumn{6}{c|}{\textbf{Normal scenario}} & \multicolumn{6}{c|}{\textbf{Attack scenario}} \\ \hline
 		Simulation time (minutes)$\,\to\,$ & \textbf{5} & \textbf{10} & \textbf{15} & \textbf{20}  & \textbf{25} & \textbf{30}  & \textbf{5}  & \textbf{10}  & \textbf{15} & \textbf{20} & \textbf{25} & \textbf{30} \\ \hline
 		& 9 & 10 & 10 & 12 & 13 & 13 & 7 & 7 & 9 & 10 & 12 & 12 \\  
 		& 1 & 1 & 1 & 1 & 1 & 1 & 8 & 9 & 6 & 7 & 12 & 13 \\  
 		& 3 & 7 & 9 & 9 & 11 & 12 & 6 & 9 & 2 & 2 & 3 & 3 \\ 
 		& 6 & 8 & 9 & 10 & 10 & 10 & 1 & 1 & 9 & 9 & 11 & 11 \\  
 		& 5 & 7 & 7 & 8 & 9 & 9 & 4 & 4 & 7 & 8 & 9 & 9 \\  
 		& 1 & 1 & 1 & 2 & 2 & 3 & 2 & 2 & {\color[HTML]{FE0000} \mybox[fill=yellow!50]{711}} & {\color[HTML]{FE0000} \mybox[fill=yellow!50]{980}} & {\color[HTML]{FE0000} \mybox[fill=yellow!50]{1246}} & {\color[HTML]{FE0000} \mybox[fill=yellow!50]{1520}} \\ 
 		&  & 2 & 3 & 3 & 4 & 5 & {\color[HTML]{FE0000} \mybox[fill=yellow!50]{166}} & {\color[HTML]{FE0000} \mybox[fill=yellow!50]{398}} & 3 & 4 & 4 & 5 \\ 
 		\multirow{-8}{*}{\begin{tabular}[c]{@{}c@{}}\rotatebox{90}{~~~DIO's received}  \rotatebox{90}{~~~~~~~~~from}  \rotatebox{90}{different neighbors}\end{tabular}} &  &  & 1 & 1 & 1 & 1 &  &  & 1 & 1 & 2 & 2 \\ \hline
 		%		\multicolumn{1}{|l|}{} &  &  &  &  &  &  &  &  &  &  &  &  \\ \hline
 		\multicolumn{1}{|c|}{ $ \tilde{x} $} & 4 & 7 & 5 & 5 & 6.5 & 7 & 6 & 7 & 6.5 & 7.5 & 10 & 10 \\ \hline
 		\multicolumn{1}{|c|}{$ Q1 $} & 1 & 1 & 1 & 1.5 & 1.5 & 2 & 2 & 2 & 2.5 & 3 & 3.5 & 4 \\ \hline
 		\multicolumn{1}{|c|}{$ Q3 $} & 6 & 8 & 9 & 9.5 & 10.5 & 11 & 8 & 9 & 9 & 9.5 & 12 & 12.5 \\ \hline
 		\multicolumn{1}{|c|}{IQR} & 5 & 7 & 8 & 8 & 9 & 9 & 6 & 7 & 6.5 & 6.5 & 8.5 & 8.5 \\ \hline
 		\multicolumn{1}{|c|}{\textit{Upper limit}} & 11 & 15 & 17 & 17.5 & 19.5 & 20 & 14 & 16 & 15.5 & 16 & 20.5 & 21 \\ \hline
 		%		\multicolumn{1}{|c|}{\textit{Lower limit}} & -4 & -6 & -7 & -6.5 & -7.5 & -7 & -4 & -5 & -4 & -3.5 & -5 & -4.5 \\ \hline
 		\multicolumn{1}{|c|}{DIO's received \textgreater \textit{Upper limit}} & $ \times $ & $ \times $ & $ \times $ & $ \times $ & $ \times $ & $ \times $ & $ \checkmark $ & $ \checkmark $ & $ \checkmark $ & $ \checkmark $ & $ \checkmark $ & $ \checkmark $\\ \hline
 		
 	\end{tabular}
 \end{table}

\section{Proposed Solution}\label{Section:ProposedSolution}
To detect \textcolor{black}{non-spoofed} copycat attack in RPL based 6LoWPANs, an IDS named as CoSec-RPL is proposed. The initial idea is to find the nodes which show significantly different behavior. CoSec-RPL is based on the idea of OD, which is also based on the IQR classifier. As discussed in section \ref{Section:Outlier detection}, the statistical method like IQR can detect outliers present in the given data with less implementation complexity. In wireless networks, the node showing abnormal behavior can be assumed as an outlier node. The existing Eq. \ref{LL} introduces a longer delay in attack detection, i.e., an attack is detected after a long time. Certain modifications have been made in the IQR method to make it fit for the detection of copycat attack in RPL based 6LoWPANs. To do this modification, a number of simulations are performed to decide the suitable value of \textit{Upper limit}. The choice of \textit{Upper limit} is based on improving the attack detection time.  Eq. \ref{LL} is modified and shown in Eq. \ref{U-UL}. The tuning parameter ($ \delta $) is responsible for improving the responsiveness of CoSec-RPL, and its value is set to 1. This has been done in order to tune the outlier detection mechanism for quick detection of the attack.

\begin{table}[!h]
	\centering
	\caption{\textcolor{black}{Network setup details of the experiments}}
	\label{tab:exp_setup_details}
	\color{black}\begin{tabular}{|c|c|c|c|c|}
		\hline
		\textbf{Scenario} & \textbf{\begin{tabular}[c]{@{}c@{}}Experiment \\ no.\end{tabular}} & \textbf{\begin{tabular}[c]{@{}c@{}}Normal \\ nodes\end{tabular}} & \textbf{\begin{tabular}[c]{@{}c@{}}Attacker\\ nodes\end{tabular}} & \textbf{Topology} \\ \hline
		\multirow{5}{*}{\textbf{\begin{tabular}[c]{@{}c@{}}\\ \\\\Normal scenario \\ (1 data packet  per \\ 60 second)\end{tabular}}} & 1 & 4 & 0 & \begin{tabular}[c]{@{}c@{}}50 m $ \times $ 50 m \\ (Random)\end{tabular} \\ \cline{2-5} 
		& 2 & 8 & 0 & \begin{tabular}[c]{@{}c@{}}50 m $ \times $ 50 m \\ (Random)\end{tabular} \\ \cline{2-5} 
		& 3 & 12 & 0 & \begin{tabular}[c]{@{}c@{}}100 m $ \times $ 100 m \\ (Random)\end{tabular} \\ \cline{2-5} 
		& 4 & 16 & 0 & \begin{tabular}[c]{@{}c@{}}100 m $ \times $ 100 m \\ (Random)\end{tabular} \\ \cline{2-5} 
		& 5 & 18 & 0 & \begin{tabular}[c]{@{}c@{}}150 m $ \times $ 150 m \\ (Random)\end{tabular} \\ \hline
		\multirow{5}{*}{\textbf{\begin{tabular}[c]{@{}c@{}}\\ \\Attack scenario \\ (Non-spoofed) \\ (1 data packet  per \\ 60 second)\end{tabular}}} & 1 & 4 & 1 & \begin{tabular}[c]{@{}c@{}}50 m $ \times $ 50 m \\ (Random)\end{tabular} \\ \cline{2-5} 
		& 2 & 8 & 1,2& \begin{tabular}[c]{@{}c@{}}50 m $ \times $ 50 m \\ (Random)\end{tabular} \\ \cline{2-5} 
		& 3 & 12 & 1,2,3 & \begin{tabular}[c]{@{}c@{}}100 m $ \times $ 100 m \\ (Random)\end{tabular} \\ \cline{2-5} 
		& 4 & 16 & 1,2,3,4 & \begin{tabular}[c]{@{}c@{}}100 m $ \times $ 100 m \\ (Random)\end{tabular} \\ \cline{2-5} 
		& 5 & 18 & 1,2,3,4 & \begin{tabular}[c]{@{}c@{}}150 m $ \times $ 150 m \\ (Random)\end{tabular} \\ \hline
	\end{tabular}
\end{table}
  
\begin{equation}
Upper~limit = Q3 + (\delta \times IQR)
\label{U-UL}
\end{equation} 

We performed multiple experiments to analyze the behavior of the network (\textcolor{black}{normal and non-spoofed attack scenarios}) in terms of the number of control messages sent and received by the nodes. \textcolor{black}{The network setup details of the experiments are shown in Table \ref{tab:exp_setup_details}} From the experiments, we observed that the node receives an almost similar number of DIO messages from its various neighbors under the normal scenario. Whereas, in the case of an attack scenario, the victim node receives a significantly large number of DIO messages from attacker node as compared to other neighbors. This makes it possible to utilize OD for detecting nodes that show abnormal behavior during network run-time. An example of OD using the modified IQR method is shown in Table \ref{tab:IQR_example}. It shows a set consisting of the number of DIO's received from different neighbors at different time intervals. The values of $ \tilde{x} $, $ Q1 $, $ Q3 $, \textit{IQR}, \textit{Upper limit} with respect to each set are tabulated. The \textit{Upper limit} acts as the safe threshold for the number of DIO received from a neighbor. The DIO count greater than the \textit{Upper limit} signifies that the respective neighbor is a copycat attacker, i.e., represented by DIO's received \textgreater \textit{Upper limit} condition. In a normal scenario, the DIO's received from each neighbor are below \textit{Upper limit}. Hence, no outlier is detected. Whereas, in the case of attack scenario, there is one neighbor from which a significantly large number of DIO's are received. Thus, the abnormal neighbor is marked as an outlier and identified as a possible copycat attacker. This detection logic has been incorporated in the CoSec-RPL for the effective detection of copycat attackers present in the network. \textcolor{black}{CoSec-RPL consists of five procedures: \textit{CoSec-RPL}, \textit{init\_neighbor\_table}, \textit{init\_blacklist\_table}, \textit{check\_malicious}, \textit{remove\_neighbor\_table\_entry}. Pseudo-codes of listed procedures are shown in Algorithm \ref{Algorithm1}, \ref{Procedure1}, \ref{Procedure2}, \ref{Procedure3}, and \ref{Procedure4}. Table \ref{tab:symbols} presents different symbols (data structures, variables) and corresponding definitions used in the proposed IDS.}

\begin{table}[!h]
	\centering
	\caption{Symbols and Definitions}
	\begin{tabular}{|c|p{8cm}|}
		\hline
		\textbf{Symbol} & \textbf{Definition}\\ \hline
		\textit{Node$_{max}$}& Maximum number of nodes in the network. \\ \hline
		\textit{$\mathcal{Z}$ $ \gets $ $ \textit{[}1,\dots,Node_{max}\textit{]} $} & Blacklist table \\ \hline
		\textit{$\mathcal{Q}$ $ \gets $ $ \textit{[}1,\dots,Node_{max}\textit{]}  $ } & Neighbor table \\ \hline
		\begin{tabular}[c]{@{}c@{}}\\ \textit{$\aleph_i$ $ \gets $ \textit{[}$ < $$ bl_{src_{ip}}$, detection$ _{count} $, status $ > $\textit{]},} \\{$i  = $ $ 1 $,\ldots, $Node_{max}$}\end{tabular} & Structure of a blacklisted node entry in blacklist table. Where, $ bl_{src_{ip}}$ represents the blacklisted node IP address, $detection_{count} $ represents the total number of times node has been detected as attacker, and $ status $ represents the status of blacklisted node, i.e., set as FALSE for suspected and TRUE for permanently blocked. \\ \hline
		\begin{tabular}[c]{@{}c@{}}\\ \textit{$\Upsilon_i$ $ \gets $ \textit{[}$ < $ from, t$ _{previous} $, t$ _{recent} $, DIO$ _{count} $$ > $\textit{]},} \\{$i  = $ $ 1 $,\ldots, $Node_{max}$}\end{tabular} & Structure of a node entry in neighbor table. Where, \textit{from} represents the DIO sender IP address, $ t_{previous} $ represents the time of previous DIO receiving, $ t_{recent} $ represents the time of most recent DIO receiving, and $ DIO_{count} $ represents the total number of DIO's received from that neighbor till current time. \\ \hline
		\textit{N$_{blacklist}$} & Number of blacklisted nodes. \\ \hline
		\textit{T$_{nodes}$} & Counter that represents total entries in neighbor table. \\ \hline
		\textit{T$ _{empty}$} & Flag to check if neighbor table and blacklist table is initialized or not.\\ \hline
		\textit{$\lambda_{current}$} & Current system clock time. \\ \hline
		\textit{ $\Psi$ } & Flag to check if the node is present in neighbor table or not.\\ \hline
		\textit{ $\rho$ } & Flag to check if the node is present in blacklist table or not.\\ \hline
		\textit{ src$ _{ip} $} & Source IP address of DIO sender node. \\ \hline
		%		\textit{ dst$ _{ip} $ }& Destination IP address \\ \hline
		$ \tau $ & Null IP address \\ \hline
		\textit{ $ \sigma $}& Safe DIO interval\\ \hline
		\textit{$\beta$} & Block threshold \\ \hline
		\textit{l} & Length of the node table at that time. \\ \hline
		\textit{active} & It indicates that IDS's detection procedure is ready to check for attackers present in neighbor table, it is set TRUE by the legitimate node after every 30 second. \\ \hline
		$ \delta $  & Tuning parameter\\ \hline
		$ \tilde{x} $ & Median \\ \hline
		$ Q1 $ &  First quartile \\ \hline
		$ Q3 $  &  Third quartile \\ \hline
		$ IQR $  & Interquartile range\\ \hline
		\textit{Upper\_limit}  & It represents the safe threshold for the number of DIO received from a neighbor. \\ \hline 
		
	\end{tabular} \label{tab:symbols}
\end{table}

\subsection{Description of \textit{CoSec-RPL} procedure}\label{Section:DescriptionOfCoSec-RPL}
\textcolor{black}{Pseudo-code of \textit{CoSec-RPL} procedure is presented in Algorithm \ref{Algorithm1}. The \textit{CoSec-RPL} procedure is incorporated in the DIO processing method, which is executed after the reception of the DIO message from any neighbor.} DIO processing method is responsible for the processing of incoming DIO messages, and executes corresponding routing management operations. \textit{CoSec-RPL} is executed every time when a DIO message is received from any neighbor.  We have considered two thresholds $ \sigma, ~\beta $, which correspond to safe DIO interval and block threshold, respectively. In addition, a tuning parameter $ \delta $ is used to control the re-activeness of \textit{CoSec-RPL}. Monitoring the time difference between successive DIO messages helps in the detection of copycat attacks. When the time difference between successive DIO messages is less than or equal to $ \sigma $, the neighbor is suspected as malicious, and vice-versa. The value of $ \sigma $ (i.e, 500 milliseconds) is adopted from Thulasiraman \textit{et al.} \cite{thulasiraman2019lightweight}. Block threshold $ \beta $ is used to avoid the permanent blocking of wrongly detected neighbors. Thus, when a neighbor is detected as an attacker, it is put in a suspected state and allowed to communicate until the block threshold is reached. Once the block threshold is reached, the neighbor is permanently blocked. The value of $ \beta $ is set to $ 5 $. \textcolor{black}{One important advantage of using $ \sigma $ in CoSec-RPL is that it helps to detect aggressive attackers which are transmitting with fixed or random intervals.} 

\textcolor{black}{This procedure is responsible for performing the following functions:}

\begin{itemize}
	\item \textcolor{black}{Initialization of neighbor and blacklist tables.}
	\item \textcolor{black}{Perform early detection of blacklisted nodes in order to minimize computational overhead.}
	\item \textcolor{black}{Maintenance of neighbor table entries, i.e., addition of new entry and updation of old entries.} 
	\item \textcolor{black}{Execution of \textit{check\_malicious} procedure after every 30 seconds to find malicious neighbors present in neighbor table.} 
\end{itemize}

\begin{algorithm}[tbh]
	\caption{Pseudo-code of CoSec-RPL}
	\label{Algorithm1}
	\begin{algorithmic}[1]
		\State \textit{Node$_{max}$, N$_{blacklist}$, T$ _{nodes} $, $\mathcal{Z}$ $ \gets $ $ \textit{[}1,\dots,Node_{max}\textit{]} $, $\mathcal{Q}$ $ \gets $ $ \textit{[}1,\dots,Node_{max}\textit{]}  $, $\lambda_{current}$, $\sigma$, $\beta$, $ \delta $} 
		\State $ \tau \gets null $, $T_{empty} \gets FALSE $ 
		\State $\aleph_i$ $ \gets $ \textit{[}$ < $$ bl_{src_{ip}}$, detection$ _{count} $, $ status $ $ > $\textit{]}, $i  = $ $ 1 $,\ldots, $Node_{max}$   
		\State \textit{$\Upsilon_i$ $ \gets $ [$ < $ from, t$ _{previous} $, t$ _{recent} $, DIO$ _{count} $ $ > $], i $ = $ $ 1 $,\ldots, Node$_{max}$} 
		\Procedure{\textit{CoSec-RPL()}}{}
		\State \textit{$\lambda_{current}$ $ \gets $ systemTime()} \Comment Get system clock time
		\If {\textit{T$ _{empty}$ $ = $ $ FALSE $}}  
		\State \textit{T$ _{empty}$$ \gets TRUE $}
		\State \textit{call init\_neighbor\_table procedure}
		\State \textit{call init\_blacklist\_table procedure}
		\EndIf
		\For{\textit{i $ \gets $ $ 1 $, N$_{blacklist}$}}
		\If{$\mathcal{Z}\textit{[}\aleph_i.bl_{src_{ip}}\textit{]}$ $ = $ $src _{ip} $ \textit{AND} $\mathcal{Z}\textit{[}\aleph_i.status\textit{]}$ $ = $ $TRUE$} \Comment Early detection
		\State \textit{discard the DIO message}
		\State \textit{\textbf{return}} 
		\EndIf
		\EndFor
		\State  $\Psi \gets FALSE$
		\For {\textit{i $ \gets $ $ 1 $, T$ _{nodes} $}} 
		\If{\textit{$\mathcal{Q}\textit{[}\Upsilon_i.from\textit{]}$ $ = $ src$ _{ip} $}}
		\State  $\Psi \gets TRUE$
		\State$\mathcal{Q}\textit{[}\Upsilon_i.t_{previous}\textit{]}$ $ \gets $ $\mathcal{Q}\textit{[}\Upsilon_i.t_{recent}\textit{]}$ 
		%		\State $\mathcal{Q}\textit{[}\Upsilon_i.t_{previous}\textit{]}$ $ \gets $ $ \lambda_{previous} $
		\State $\mathcal{Q}\textit{[}\Upsilon_i.t_{recent}\textit{]}$ $ \gets $ $\lambda_{current}$ 
		\State $\mathcal{Q}\textit{[}\Upsilon_i.DIO_{count}\textit{]}$ $ \gets $ $\mathcal{Q}\textit{[}\Upsilon_i.DIO_{count}\textit{]}$ + $ 1 $     
		\EndIf
		\EndFor
		\If{\textit{$\Psi$ $ = $ FALSE}}
		\For{\textit{i $ \gets $ 1, Node$_{max}$}}
		\If{\textit{$\mathcal{Q}\textit{[}\Upsilon_i.from\textit{]}$ $ = $ $\tau$}}
		\State$\mathcal{Q}\textit{[}\Upsilon_i.t_{previous}\textit{]}$ $ \gets $ $\mathcal{Q}\textit{[}\Upsilon_i.t_{recent}\textit{]}$ 
		%		\State $\mathcal{Q}\textit{[}\Upsilon_i.t_{previous}\textit{]}$ $ \gets $ $ \lambda_{previous} $ 
		\State $\mathcal{Q}\textit{[}\Upsilon_i.t_{recent}\textit{]}$ $ \gets $ $\lambda_{current}$ 
		\State $\mathcal{Q}\textit{[}\Upsilon_i.DIO_{count}\textit{]}$ $ \gets $ $\mathcal{Q}\textit{[}\Upsilon_i.DIO_{count}\textit{]}$ + $1 $
		\State $T_{nodes}$++  
		\EndIf
		\EndFor
		\EndIf
		\If{$ active = TRUE $}
		\State \textit{call check\_malicious procedure} \Comment Check for malicious neighbors in neighbor table
		\State $ active \gets FALSE $
		\EndIf
		\EndProcedure

	\end{algorithmic}
\end{algorithm}        
            
\subsection{Description of \textit{check\_malicious} procedure}\label{Section:Description of check_malicious procedure}
The \textit{check\_malicious} procedure is the most important part of the CoSec-RPL scheme. \textcolor{black}{It is responsible for finding the malicious neighbors (i.e., outliers) present in the neighbor table. Pseudo-code of \textit{check\_malicious} is presented in Algorithm \ref{Procedure1}. The count of DIO messages received by different neighbors is used to filter out the malicious nodes. Also, the modified IQR method is implemented to compute the safe threshold (i.e., \textit{Upper\_limit}). Neighbors are checked against \textit{Upper\_limit} and safe DIO interval. In case any neighbor violates the safety conditions, it is marked as suspected and added to the blacklist table. A node is marked malicious and permanently blocked if it is suspected for $ \beta $ times. Upon detection of malicious neighbor its entry is removed from neighbor table.}

\floatname{algorithm}{Algorithm}
\begin{algorithm}[tbh]
	\caption{Pseudo-code of check\_malicious procedure}
	\label{Procedure1}
	\begin{algorithmic}[1]
		\Procedure{\textit{check\_malicious()}}{} \Comment Checks for malicious neighbors present in neighbor table.   
		\State $ l \gets 0 $ \Comment Variable to store current  length of neighbor table
		\For{\textit{i $ \gets $ $ 1 $, Node$_{max}$}}
		\If{\textit{$\mathcal{Q}\textit{[}\Upsilon_i.from\textit{]}$ $ != $ $ \tau $}}
		\State $ l $++
		\EndIf
		\EndFor
		%		\State \textit{call COMPUTE\_INDEXES (l) procedure}
		\If {$ l > 1 $}
		\State \textit{sort} $\mathcal{Q}$ \textit{on} $ DIO_{count} $ \textit{column}
		\EndIf 
		\State \textit{compute} $ \tilde{x},~Q1,~Q3 $ \textit{values of} $\mathcal{Q}\textit{[}\Upsilon_i.DIO_{count}\textit{]}$, \textit{where} $i =  1,\ldots, l$
		\State $ IQR \gets Q3-Q1 $  
		\State \textit{Upper\_limit} $\gets Q3 + (\delta \times IQR) $
		\For {\textit{i $ \gets $ $ 1 $, l}}
		\If{$\mathcal{Q}\textit{[}\Upsilon_i.DIO_{count}\textit{]} > Upper\_limit$}
		\If{$\mathcal{Q}\textit{[}\Upsilon_i.t_{recent}\textit{]}-\mathcal{Q}\textit{[}\Upsilon_i.t_{previous}\textit{]}\leq \sigma $}
		\State $ \rho \gets FALSE  $
		\For {$j  \gets  1, N_{blacklist}$}
		\If{$\mathcal{Z}\textit{[}\aleph_j.bl_{src_{ip}}\textit{]} = \mathcal{Q}\textit{[}\Upsilon_i.from \textit{]}$}
		\State $ \rho \gets TRUE  $ 
		\If{$\mathcal{Z}\textit{[}\aleph_j.detection_{count}\textit{]}<\beta$}
		\State$\mathcal{Z}\textit{[}\aleph_j.detection_{count}\textit{]} \gets \mathcal{Z}\textit{[}\aleph_j.detection_{count}\textit{]} + 1 $
		\If{$ \mathcal{Z}\textit{[}\aleph_j.detection_{count}\textit{]} = \beta $}
		\State $ \mathcal{Z}\textit{[}\aleph_j.status\textit{]} \gets TRUE$
		\State \textit{Neighbor is permanently blocked}
		\State \textit{call remove\_neighbor\_table\_entry(i) procedure} 
		\EndIf
		\EndIf
		\EndIf
		\EndFor
		\If{$ \rho = FALSE $}
		\State$  k \gets  N_{blacklist}++ $
		\State $\mathcal{Z}\textit{[}\aleph_k.bl_{src_{ip}}\textit{]} \gets \mathcal{Q}\textit{[}\Upsilon_i.from \textit{]} $ 
		\State $ \mathcal{Z}\textit{[}\aleph_k.detection_{count}\textit{]} \gets \mathcal{Z}\textit{[}\aleph_k.detection_{count}\textit{]}+ 1$
		\State $ \mathcal{Z}\textit{[}\aleph_k.status\textit{]} \gets FALSE$
		\State \textit{Neighbor is suspected to be an attacker}
		\EndIf
		\EndIf
		\EndIf   
		\EndFor
		\EndProcedure
		
	\end{algorithmic}
\end{algorithm}

\subsection{Description of \textit{init\_neighbor\_table, init\_blacklist\_table, remove\_neighbor\_table\_entry} procedures}\label{Section:DescriptionOfSuppelmentry procedures}

\textcolor{black}{The \textit{init\_neighbor\_table, init\_blacklist\_table, remove\_neighbor\_table\_entry} are supporting procedures of CoSec-RPL scheme. The \textit{init\_neighbor\_table} procedure initializes the neighbor table when node is powered ON. Pseudo-code of \textit{init\_neighbor\_table} is shown in \textcolor{black}{Algorithm} \ref{Procedure2}. The \textit{init\_neighbor\_table} procedure is responsible for initializing the neighbor table entries with default values. Pseudo-code of \textit{init\_blacklist\_table} is presented in \textcolor{black}{Algorithm} \ref{Procedure3}. The \textit{init\_blacklist\_table} procedure initializes the blacklist table when node is powered ON.  It initializes the blacklist table entries with default values. Pseudo-code of \textit{remove\_neighbor\_table\_entry} is illustrated in \textcolor{black}{Algorithm} \ref{Procedure4}. The \textit{remove\_neighbor\_table\_entry} procedure deletes the neighbor table entry.}

\floatname{algorithm}{Algorithm}
\begin{algorithm}[tbh]
	\caption{Pseudo-code of init\_neighbor\_table procedure}
	\label{Procedure2}
	\begin{algorithmic}[1]
		\Procedure{\textit{init\_neighbor\_table()}}{} \Comment Initializes neighbor table entries.   
		%		\If{\textit{NETSTACK\_CONF\_WITH\_IPV6}}
%		\State  \textit{$ \tau $  $ \gets $  $ null $} \Comment Null IP address for IPv6 network
		\For{\textit{i $ \gets $ $ 1 $, Node$_{max}$}}
		\State \textit{$\mathcal{Q}\textit{[}\Upsilon_i.from \textit{]}$ $ \gets $ $ \tau $}
		\State $\mathcal{Q}\textit{[}\Upsilon_i.t_{previous}\textit{]}$ $ \gets $ 0
		\State $\mathcal{Q}\textit{[}\Upsilon_i.t_{recent}\textit{]}$ $ \gets $ 0 
		\State $\mathcal{Q}\textit{[}\Upsilon_i.DIO_{count}\textit{]}$ $ \gets $ 0
		\EndFor
		\EndProcedure
		
	\end{algorithmic}
\end{algorithm}

\floatname{algorithm}{Algorithm}
\begin{algorithm}[tbh]
	\caption{Pseudo-code of init\_blacklist\_table procedure}
	\label{Procedure3}
	\begin{algorithmic}[1]
		\Procedure{\textit{init\_blacklist\_table()}}{} \Comment Initializes blacklist table entries.   
		%		\If{\textit{NETSTACK\_CONF\_WITH\_IPV6}}
%		\State \textit{$ \tau $  $ \gets $  $ null $} \Comment Null IP address for IPv6 network
		\For{\textit{i $ \gets $ $ 1 $, Node$_{max}$}}
		\State $\mathcal{Z}\textit{[}\aleph_i.bl_{src_{ip}}\textit{]}$ $ \gets $ $ \tau $
%		\State $\mathcal{Z}$$\textit{[i]}.bl_{src_{ip}}$ $ \gets $ $ \tau $
		\State $\mathcal{Z}\textit{[}\aleph_i.detection_{count}\textit{]}$ $ \gets $ $ 0 $
%		\State $\mathcal{Z}$$\textit{[i]}$$.detection_{count}$ $ \gets $ $ 0$
		\State $\mathcal{Z}\textit{[}\aleph_i.status\textit{]}$ $ \gets $ $ FALSE $
%		\State $\mathcal{Z}$$\textit{[i]}$$.status$ $ \gets $ $ FALSE$  
		\EndFor
		\EndProcedure
		
	\end{algorithmic}
\end{algorithm}

\floatname{algorithm}{Algorithm}
\begin{algorithm}[tbh]
	\caption{Pseudo-code of remove\_neighbor\_table\_entry procedure}
	\label{Procedure4}
	\begin{algorithmic}[1]
		\Procedure{\textit{remove\_neighbor\_table\_entry}}{\textit{location}} \Comment Removes neighbor table entry.   
		\State \textit{$\mathcal{Q}\textit{[}\Upsilon_{location}.from \textit{]}$ $ \gets $ $ \tau$}
		\State  $\mathcal{Q}\textit{[}\Upsilon_{location}.t_{previous}\textit{]}$ $ \gets $ 0
		\State $\mathcal{Q}\textit{[}\Upsilon_{location}.t_{recent}\textit{]}$ $ \gets $ 0 
		\State $\mathcal{Q}\textit{[}\Upsilon_{location}.DIO_{count}\textit{]}$ $ \gets $ 0
		\EndProcedure
		
	\end{algorithmic}
\end{algorithm}

\section{Performance Evaluation}\label{Section:PerformanceEval}
In this section, we first focus on studying the impact of the copycat attack on the network's performance. Then a detailed evaluation of the proposed CoSec-RPL scheme is done.  \textcolor{black}{A number of experiments have been performed using the Cooja simulator, which is the most reliable and widely used network simulator provided in Contiki operating system \cite{dunkels2011contiki}.} Contiki is a well known lightweight and publicly available operating system for constrained devices.    

\subsection{Experimental setup}\label{Section:ExperimentalSetup}
\textcolor{black}{Cooja is capable of producing real results for evaluations. It has an inbuilt hardware simulator named MSPsim that emulates the exact binary code of real sensor devices in order to achieve realistic simulation.} In this paper, Zolertia 1 (Z1) platform is utilized to act as a 6LoWPAN node. Table \ref{table1} presents the simulation parameters considered in the experiments. \textcolor{black}{In order to simulate a realistic scenario, the Multipath Ray-Tracer Medium (MRM) radio model is used in all the experiments \cite{Perazzo2017,wang2017routing, ancillotti2018rtt, kanaris2019realistic}}. The MRM radio model parameters shown in Table \ref{table2} have been adopted from Perazzo \textit{et} al. \cite{Perazzo2017, vallatiemail}. \textcolor{black}{A network topology of 16 sensors randomly distributed in a square grid of 150 m $ \times $ 150 m resembles smart agriculture and small industry monitoring application. A small level deployment of these applications involves the placement of several monitoring nodes that cover farmland or industrial place. Thus the considered network settings are sufficient for LLN security study.} All the nodes are running on Contiki with a common protocol stack, as shown in Table \ref{table3}. The ContikiRPL library is modified to implement the copycat attack on attacker nodes as well as to implement the proposed IDS on legitimate sensor nodes. Specifically, an attacker node is programmed to eavesdrop and capture the DIO message from any legitimate node and then replay the captured message at a fixed replay interval. A network scenario containing 1 gateway node and 16 sensor nodes which are randomly placed on a grid of 150m$\times$150m is considered. Each sensor sends a data packet (30 bytes) to a gateway every 60 seconds. Random Waypoint Mobility Model is used to simulate the behavior of mobile nodes where the speed of nodes is set between $ 1 $ m/s and $ 2 $ m/s \cite{kabilan2018performance}. In order to perform fair experiments, the attacker node is programmed to launch an attack after 90 seconds of network initialization. In this way, the attack starts after the network is established and becomes stable. Similarly, CoSec-RPL is programmed to activate after 120 seconds of network initialization and repeatedly checks for malicious neighbors every 30 seconds.  

%We assume that the network is static, and the attacker has compromised some legitimate internal nodes and reprogrammed them to perform a copycat attack (insider attack scenario).       

\begin{table}[!h]
	\centering
	\caption{Simulation parameters}
	\begin{tabular}{|l|p{4.5cm}|}
		\hline
		\textbf{Parameter} & \textbf{Values}\\ \hline
		Radio model & Multipath Ray-Tracer Medium (MRM) \\ \hline
		Mobility model & Random Waypoint Mobility Model \\ \hline
		Simulation area & 150 m $ \times $ 150 m\\ \hline
		Simulation time & 1800 seconds \\ \hline
		Objective function & Minimum Rank with Hysteresis Objective Function(MRHOF) \\ \hline
		Number of attacker nodes & 4 \\ \hline
		Number of gateway nodes & 1 \\ \hline
		Number of sensor nodes & 16 \\ \hline
		DIO minimum interval & 4 seconds \\ \hline
		DIO maximum interval & 17.5 minutes \\ \hline
		Replay interval & 1, 2, 3, 4 seconds \\ \hline
		Data packet size & 30 bytes \\\hline
		Data packet sending interval & 60 seconds \\ \hline
		Transmission power & 0 dBm \\\hline
		Node speed & $ 1 $ m/s$-$$ 2 $ m/s \\\hline
		
	\end{tabular} \label{table1}
\end{table}

\begin{table}[!h]
	\centering
	\caption{MRM parameters}
	\begin{tabular}{|c|c|}
		\hline
		\textbf{Parameter} & \textbf{Value} \\ \hline
		tx\_power & 0.0 \\ \hline
		tx\_with\_gain & false \\ \hline
		captureEffect & false \\ \hline
		obstacle\_attenuation & -10.0 \\ \hline
		system\_gain\_mean & -20.0 \\ \hline
		system\_gain & 0.0 \\ \hline
	\end{tabular}

	\label{table2}

\end{table}	

\begin{table}[!h]
	\centering
	\caption{Contiki parameters}
\begin{tabular}{|p{4.3cm}|p{2.1cm}|p{4.5cm}|}
		\hline
		\textbf{Parameter} & \textbf{Value}&  \textbf{Description}\\ \hline
		NETSTACK\_CONF\_WITH\_IPV$ 6 $ & 1 & Configured to enable IPv$ 6 $ networking.\\ \hline
		NETSTACK\_CONF\_NETWORK & sicslopan\_driver & Enables header compression and fragmentation.\\ \hline
		NETSTACK\_CONF\_MAC &  csma\_driver& Enables Media Access Control with Collision Avoidance.\\ \hline
		NETSTACK\_CONF\_RDC &  contikimac\_driver & Enables energy efficiency using radio duty cycling (RDC) \\ \hline
		NETSTACK\_CONF\_RADIO & cc$ 2420 $\_driver & Control the operation of IEEE $ 802.15.4 $ compliant CC$ 2420 $ radio transceiver operating at $ 2.4 $ Ghz.\\ \hline
		NETSTACK\_CONF\_FRAMER & framer\_$ 802154 $ & Enables parsing and generation of formatted packets compatible with IEEE $ 802.15.4 $ protocol.\\ \hline
	\end{tabular} \label{table3}
\end{table}

\subsection{Performance indicators}
In order to analyze the impact of \textcolor{black}{non-spoofed} copycat attack on RPL based 6LoWPAN network, PDR and AE2ED are selected. Similarly to evaluate CoSec-RPL, Attacker Detection Accuracy (ADA) and First Response Time (FRT) are analyzed. These performance indicators are defined as, 

\begin{enumerate}
	\item \textit{PDR}: It is the ratio between the total number of data packets received by the gateway node to the total data packets sent by the sensor nodes including re-transmitted data packets. PDR is calculated as Eq. \ref{Eq-PDR}.
	
	\begin{equation}
	PDR = \frac{D_{received}}{\sum_{j=1}^{N} D_{sent_j}}
	\label{Eq-PDR}
	\end{equation} 
	where $ D_{received} $ represents the total number of data packets received at gateway node, and $ D_{sent_j} $ represents the total data packets sent from non-root node $ j $.
	
	\item \textit{AE2ED}: AE2ED is defined as the average amount of time taken by all the data packets sent from each sensor node, to be successfully delivered to the gateway node while neglecting all lost and dropped packets. AE2ED is calculated as Eq. \ref{Eq-AE2ED}.
	\begin{equation}
	AE2ED = \frac{\sum_{j=1}^{N} D_{received_j}}{D_{N}}
	\label{Eq-AE2ED} 
	\end{equation}
	where $ D_{received_{j}} $, $ D_{N} $ represent time delay of data packet $ j $ and total number of received packets, respectively.
	
	\item  \textit{ADA}: It represents the ratio of total number of correctly detected attackers with respect to all observations made by IDS. ADA is calculated as Eq. \ref{ADA}.
	\begin{equation}
	ADA = \frac{A_{True}}{A_{True}+A_{False}}
	\label{ADA}
	\end{equation}
	where $ A_{True} $, $ A_{False} $ represent correctly and wrongly detected attackers, respectively. 
		
	\item \textit{FRT}: It is defined as the time interval between the attack launch by a particular attacker and its first detection by IDS. FRT is calculated as Eq. \ref{FRT}.
	\begin{equation}
	FRT = T_{first~detection} - T_{launch}
	\label{FRT}
	\end{equation}  
	where $ T_{first~detection} $, $ T_{launch} $ represent time of first detection and time of launch, respectively.   
\end{enumerate} 

\subsection{Simulation results}
Both static and mobile (dynamic) network scenarios are studied in this paper. First, the impact of \textcolor{black}{non-spoofed} copycat is analyzed on the network in terms of PDR and AE2ED. Second, the performance of the CoSec-RPL scheme is studied in terms of ADA and FRT. For each scenario, 10 independent replications with different seeds are run in order to obtain statistically valid results. The mean values of the obtained results with its errors at 95\% confidence interval have been reported to avoid biased observations. 

\subsubsection{Impact on PDR}\label{Impact on PDR}

\begin{figure}[!h]
	\centering
	\includegraphics[width = 1.1\textwidth]{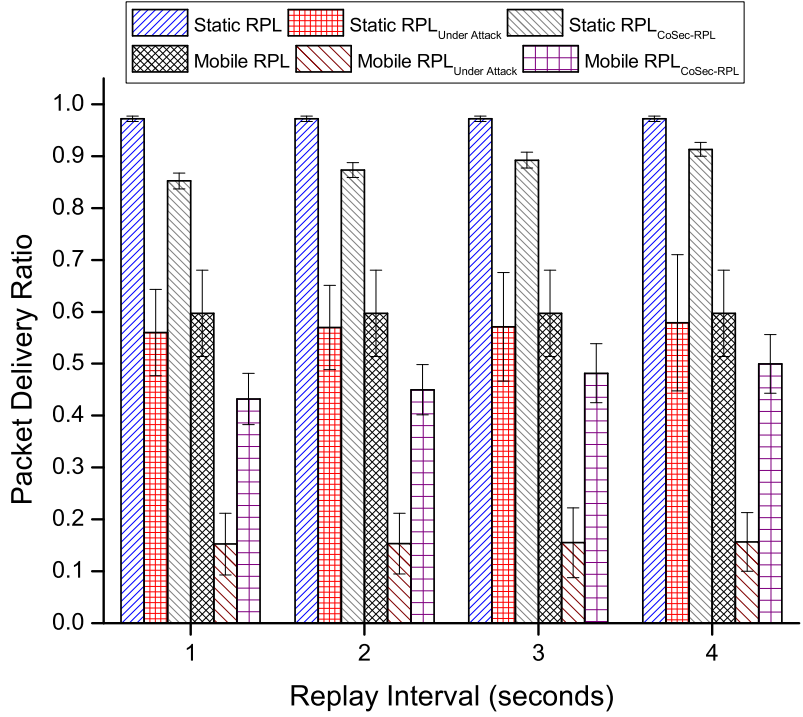}
	\caption{\textcolor{black}{PDR values obtained in different scenarios}}
	\label{PDR}
\end{figure}

\textcolor{black}{The performance of Static RPL (static network reference model without attack), Static RPL\textsubscript{Under Attack} (i.e., Static RPL under \textcolor{black}{non-spoofed} copycat attack), Static RPL\textsubscript{CoSec-RPL} (i.e., Static RPL under attack with our proposed defense scheme), Mobile RPL (mobile network reference model without attack), Mobile RPL\textsubscript{Under Attack} (i.e., Mobile RPL under \textcolor{black}{non-spoofed} copycat attack), and Mobile RPL\textsubscript{CoSec-RPL} (i.e., Mobile RPL under attack with our proposed defense scheme) is evaluated and compared.} In the case of the Static RPL and Mobile RPL, it must be noted that the replay interval plays no role. Achieving good data packet delivery is one of the major requirements of critical IoT applications. Thus, PDR analysis is an essential criterion in the performance evaluation of 6LoWPANs. Fig. \ref{PDR} shows PDR obtained with different replay intervals, i.e., $ 1, 2, 3,$ and $4 $ seconds. It can be observed that the attack severely degrades the performance of the network. This is confirmed from the comparison of PDR values of Static RPL vs. Static RPL\textsubscript{Under Attack}, and Mobile RPL vs. Mobile RPL\textsubscript{Under Attack}, under different attack intervals. 

The PDR values achieved in Static RPL, Mobile RPL are $ \approx $0.97 and $ \approx $0.59, respectively. On the other hand in Static RPL\textsubscript{Under Attack}, Mobile RPL\textsubscript{Under Attack} the PDR is reduced to $ \approx $0.57 and $ \approx $0.14, respectively. The \textcolor{black}{non-spoofed} copycat attack induces a major impact on network's PDR. The main reason for this is that during the attack, a victim node repeatedly receives DIO (with \textcolor{black}{non-spoofed} source IP address) messages from an unresponsive attacker, in very short interval. From an unresponsive attacker, we mean that the node that does not respond to the victim's DAO messages (in case downward routing is enabled). This forces the victim node to perform unnecessary routing management related operations on every illegitimate DIO reception, which limits its data packet forwarding behavior. Such a reduction in PDR is unsuitable for critical IoT applications like healthcare. Hence, \textcolor{black}{non-spoofed} copycat attack must be addressed for smooth operation of critical applications. \textcolor{black}{CoSec-RPL is able to improve the PDR of both static and mobile networks. Average PDR values achieved in case of Static RPL\textsubscript{CoSec-RPL} and Mobile RPL\textsubscript{CoSec-RPL} are $ \approx $0.88 and $ \approx $0.46, respectively. It can be observed from the PDR values achieved in the case of Static RPL\textsubscript{CoSec-RPL} and Mobile RPL\textsubscript{CoSec-RPL} that network's performance is improved. It is because CoSec-RPL detects and blocks all the incoming packets from the attacker node and consequently, reduces the effect of the attack on legitimate nodes.}

\subsubsection{Impact on AE2ED} \label{Impact on AE2ED}

\begin{figure}[!h]
	\centering
	\includegraphics[width = 1.1\textwidth]{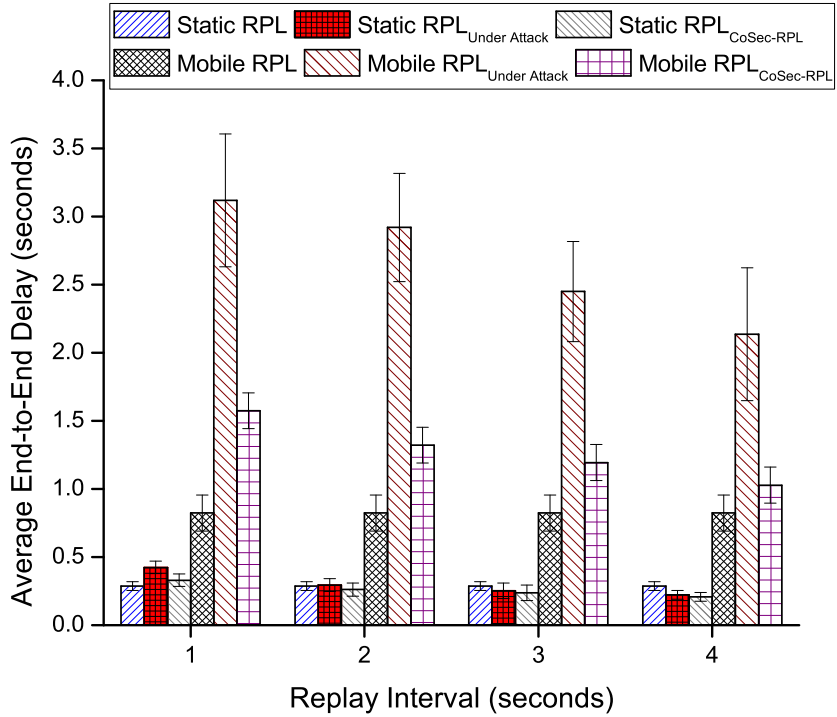}
	\caption{\textcolor{black}{AE2ED values obtained in different scenarios}}
	\label{AE2ED}
\end{figure}

There are a number of critical IoT applications that can't tolerate network latency issues. Thus, it is also essential to make sure the network has minimal latency. In this regard, the impact of \textcolor{black}{non-spoofed} copycat attack on AE2ED of 6LoWPAN network is studied. Fig. \ref{AE2ED} shows AE2ED obtained with different replay intervals. It can be observed that the attack increases network latency. This is confirmed from the comparison of AE2ED values of Static RPL vs. Static RPL\textsubscript{Under Attack}, and Mobile RPL vs. Mobile RPL\textsubscript{Under Attack}, under different attack intervals.

It can also be observed in Fig. \ref{AE2ED} that with different attack intervals, the AE2ED values obtained in Static RPL, Mobile RPL are $\approx$0.28 and $\approx$0.82, respectively. Whereas, in case of Static RPL\textsubscript{Under Attack} the AE2ED is achieved between $ \approx $0.22 and $ \approx $0.42. In case of Mobile RPL\textsubscript{Under Attack} the AE2ED is achieved between $ \approx $\textcolor{black}{2.13} and $ \approx $3.11. It can be seen that the attack does not have any significant impact on AE2ED of the static network. On the other hand, in the case of the attack on a mobile network, AE2ED significantly increases. This is because of the network dynamicity due to which the nodes frequently leave and join the DODAG. This situation consequently, leads to frequent DODAG repairs and parent switching, which provides an attacker with a major benefit to increasing the attack's impact on the network. Mobile RPL\textsubscript{Under Attack} achieves lowest and highest AE2ED with \textcolor{black}{$ 4 $} and $ 1 $ second replay interval, respectively. AE2ED of the network under attack increases because of two major reasons. The first reason is the congestion and interference evoked by the \textcolor{black}{non-spoofed} copycat attacker, which affects the forwarding nodes in the attack region. The second reason is the creation of non-optimal routes due to the replay of outdated routing information, which increases the path length for routing data packets. \textcolor{black}{CoSec-RPL improves the AE2ED of both static and mobile network scenarios. The average AE2ED values achieved in case of Static RPL\textsubscript{CoSec-RPL} and Mobile RPL\textsubscript{CoSec-RPL} are $ \approx $0.25 and $ \approx $1.27, respectively. It can be observed from the AE2ED values achieved in the case of Static RPL\textsubscript{CoSec-RPL} and Mobile RPL\textsubscript{CoSec-RPL} that network's performance is significantly improved. CoSec-RPL effectively reduces the time required for routing data packets from node to 6BR by detecting and blocking the incoming malicious traffic from the attacker node. CoSec-RPL reduces the computational overhead induced on legitimate nodes due to reception of outdated routing information from the attacker node.}

\subsubsection{IDS performance in terms of ADA}\label{Section:ADA}

\begin{figure}[!h]
	\centering
	\includegraphics[width = \textwidth]{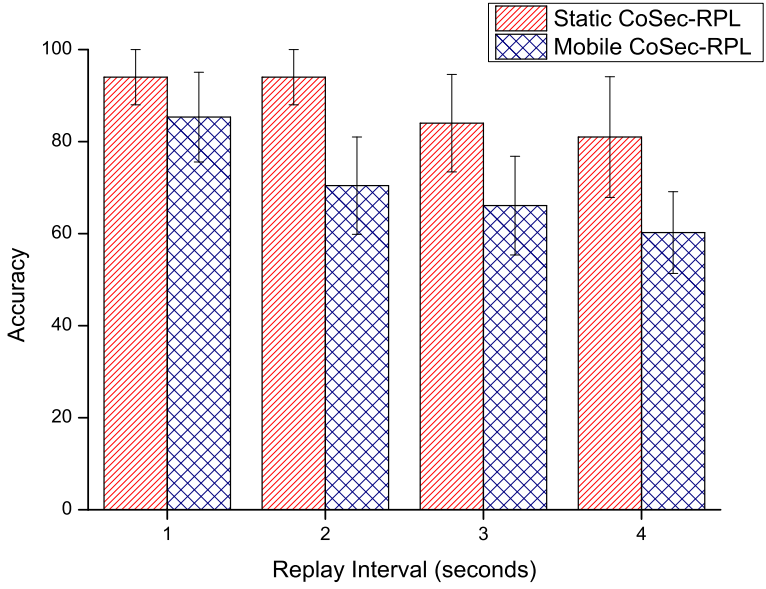}
	\caption{Detection Accuracy of proposed IDS}
	\label{fig:ADA}
\end{figure}

Fig. \ref{fig:ADA} shows the ADA achieved by CoSec-RPL in different attack scenarios. Where, Static CoSec-RPL, Mobile CoSec-RPL represent results obtained from CoSec-RPL operation in a static and mobile network scenarios, respectively. It can be observed from the results that CoSec-RPL performs better in a static network by achieving a maximum of 94\% and a minimum of 81\% ADA. Whereas, in the case of mobile network, CoSec-RPL achieves a maximum of $ \approx $85\% and a minimum $ \approx $60\% ADA. CoSec-RPL performs well in the static network because of the stable network due to which the attack detection mechanism is able to correctly identify the malicious neighbors present in the node's neighbor table. On the contrary, mobile networks are dynamic where frequent leaving and joining of legitimate nodes increases the number of DIO message transmissions. The legitimate nodes which have transmitted many DIO messages in order to join the DODAG become suspected attacker because they are detected as an outlier by CoSec-RPL's attack detection mechanism. However, the permanent blocking of legitimate nodes is still prevented because of the block threshold ($ \beta $). The performance of CoSec-RPL in terms of ADA is inversely proportional to the replay interval. This indicates that CoSec-RPL is able to detect aggressive attackers more accurately than the non-aggressive attacker. Assuming that the attacker chooses an aggressive strategy to create a major impact on the network, CoSec-RPL is the suitable choice to detect such attacks.

\subsubsection{IDS performance in terms of FRT}
\begin{figure}[!h]
	\centering
	\includegraphics[width=1.1\textwidth]{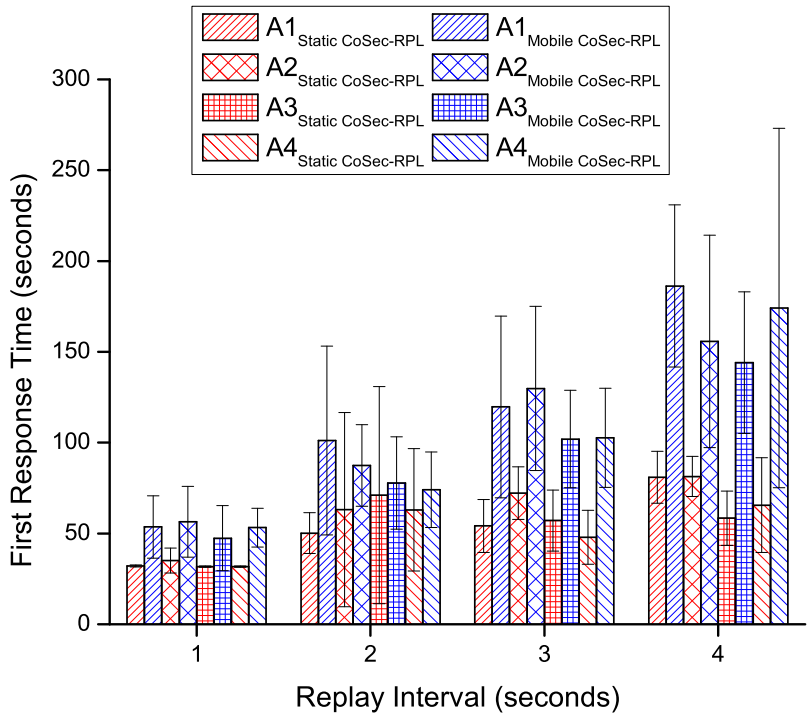}
	\caption{FRT of CoSec-RPL to detect attackers with different replay intervals}
	\label{fig:FRT}
\end{figure}

The responsiveness of an IDS plays a major role for deciding its usefulness in real-world applications. \textcolor{black}{Considering this important factor, we have analyzed the performance of CoSec-RPL in terms of FRT. Fig. \ref{fig:FRT} shows FRT of CoSec-RPL to detect attackers with different replay intervals. The results have been reported for each attacker individually, which are represented by A1, A2, A3, and A4. It can be seen that the Static CoSec-RPL performs better than Mobile CoSec-RPL.} This is because of the stable network that makes it easy for the detection mechanism to quickly find the malicious neighbor present in the neighbor table. The reason for delayed attacker detection in the case of Mobile CoSec-RPL is the network dynamicity, which increases the DIO transmission of legitimate nodes. Hence,  it becomes very difficult for the detection mechanism to differentiate between normal and attacker neighbors present in the neighbor table. As mentioned in section \ref{Section:ADA}, aggressive attacker is quickly detected by CoSec-RPL as compared to a non-aggressive attacker. \textcolor{black}{ A Similar pattern is observed in FRT results shown in Fig. \ref{fig:FRT}.} In both static and mobile scenarios, CoSec-RPL achieves minimum FRT to detect most aggressive attackers, i.e., A1-A4 with 1 second replay interval. Whereas maximum FRT is achieved in the detection of least aggressive attackers, i.e., A1-A4 with 4 second replay interval. From FRT analysis, it can be concluded that the performance of CoSec-RPL dependent on the replay interval of the attacker. Small replay interval corresponds to quick and accurate intrusion detection, while large replay interval corresponds to slower and less accurate intrusion detection.

\subsubsection{Implementation Overhead}

\begin{figure}[!h]
	\centering
	\includegraphics[width =\textwidth]{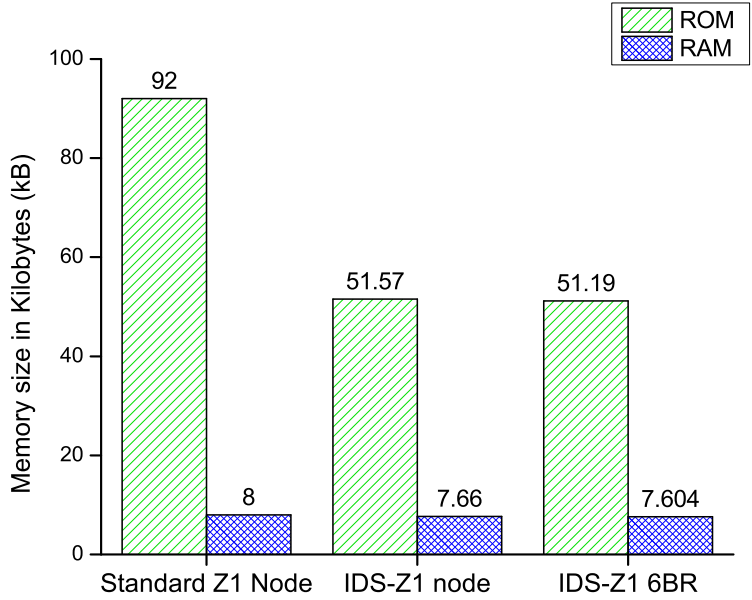}
	\caption{Memory requirements of CoSec-RPL}
	\label{Memory}
\end{figure}

The resource constrained nature of 6LoWPAN nodes restrict the usage of resource-hungry security solutions. Thus, it is very important to develop a lightweight security solution that does not consume a lot of node's resources, i.e., CPU, memory, and energy. In this section, the implementation overhead of CoSec-RPL in terms of memory requirement (i.e., static memory (RAM) and flash memory (ROM)) is analyzed and discussed. To determine the memory requirements of CoSec-RPL, the msp430-size tool is utilized. Fig. \ref{Memory} shows the comparison between the memory requirements of the sensor node and the 6BR node over which the CoSec-RPL is implemented. \textcolor{black}{On a sensor node, the Z1 binary with CoSec-RPL implemented on it requires 51.57 kB of ROM and 7.66 kB of RAM}. \textcolor{black}{On 6BR node, Z1 binary occupies 51.19 kB of ROM and 7.604 kB of RAM.} The maximum ROM, RAM storage capacity of Z1 node is 92 kB and 8 kB,  respectively. CoSec-RPL modules additionally require only 5.9 kB, 2.56 kB of ROM and RAM, respectively. This indicates that the proposed IDS can be used to secure resource constrained nodes from \textcolor{black}{non-spoofed} copycat attack.

\subsubsection{\textcolor{black}{Theoretical comparison with existing works}}
\textcolor{black}{Table \ref{tab:RelatedWork} presents the details of some RPL specific IDS proposed in the literature. Also, a theoretical comparison of recent literature is made with our proposed solution. The main motivation behind carrying out theoretical comparison is that the existing RPL specific IDS are developed to detect different (other than copycat attack) routing attacks such as DIS flooding, rank, version number, etc. These IDS are specifically designed to detect a particular type of attack and will fail to detect copycat attack. A fair comparison can only be made with those IDS which have been designed to detect the copycat attack. To the best of our knowledge, there are no IDS present in literature that has been designed to detect the copycat attack, and hence we only rely on the theoretical comparison. The same methodology for performance comparison is followed in similar existing works \cite{Raza1043301, Airehrour2018, ghaleb2018addressing, wadhaj2020mitigation, mayzaud2015mitigation, verma2020mitigation} which inspired us to carry out practical comparison with standard RPL only.}

\begin{table}[!h]
	\centering
	\caption{\textcolor{black}{Theoretical comparison of RPL specific IDS present in the literature with our proposed solution}}
	\begin{tabular}{|p{.5cm}|p{2cm}|p{1.2cm}|p{7cm}|p{2cm}|}
		\hline
		\textbf{Ref.} & \textbf{Defense Mechanism} & \textbf{Mobility support} & \textbf{Limitations} & \textbf{\textcolor{black}{Detection of non-spoofed copycat attack}} \\ \hline
		
		\cite{ghaleb2018addressing}& SecRPL &No & Degrades the network performance in terms of power consumption, control packet overhead, latency, and network reliability. & No\\ \hline
		\cite{Airehrour2018}& SecTrust-RPL &No & Requires promiscuous mode of operation for constant monitoring.  &  No\\ \hline
		\cite{RAZA20132661}&SVELTE & & Synchronization issue, requires strategic placement of IDS modules, vulnerable to coordinated attacks. &No \\ \hline
		\cite{Mayzaud2017}& Distributed Monitoring Architecture  & No & Requires promiscuous mode of operation for constant monitoring, relies on high order devices for monitoring which adds cost overhead, requires strategic placement of monitoring nodes. & No\\ \hline
		\cite{GaraInderscience} & Hybrid IDS based on Sequential Probability Ratio Test with an Adaptive Threshold & Yes & Increases network overhead due to use of HELLO messages. &  No\\ \hline
		\cite{Bostani2017} & Hybrid of Anomaly and Specification based IDS & No& Only suitable for applications with one way communication. & No\\ \hline
		\cite{Le2016} & Specification based IDS & No& Introduces communication overhead, requires a good network trace for the creation of effective specification, and shows less accuracy when it works for a long time. & No \\ \hline
		\textcolor{black}{\cite{verma2020mitigation}} & \textcolor{black}{Secure-RPL} & \textcolor{black}{Yes} & \textcolor{black}{Requires minor changes in RPL implementation, performance is dependent on proper
			selection of safety thresholds.}   & \textcolor{black}{No} \\ \hline
		- & CoSec-RPL (proposed solution) & Yes & Requires minor changes in RPL implementation, requires to maintain a neighbor table to store neighbor information.    & Yes \\ \hline
		
	\end{tabular} \label{tab:RelatedWork}
\end{table}

\section{CoSec-RPL future extensions}\label{Section:CoSec-RPL extensions}
One of the main advantages of CoSec-RPL is that it can be easily extended for the detection of other attacks. The detection mechanism of our proposed IDS can be adapted for the detection of attacks that involve an attacker to send a large number of control messages to legitimate nodes. OD mechanism can be improved by using Kalman filter (statistics and control theory) \cite{wang2018robust} and Entropy (Information theory) \cite{zhi2018gini, domingues2018comparative}.   
\\
\textit{DIS flooding attack detection}: The proposed IDS can be modified to detect the DIS flooding attack. It will only require two modifications: (1) it will require an extra filed in the neighbor table that stores the number of DIS messages received from neighbors; (2) based on the threshold on a maximum allowed DIS messages from the neighbor, DIS flooding attacker can be detected.       
\\
\textit{DAO insider attack detection:} In the DAO insider attack, an insider attacker node sends fake DAO messages to its parent repeatedly in a fixed interval. In this way, an attacker generates a flood of DAO messages. CoSec-RPL can be extended to maintain the count of DAO messages received from child nodes. The node having abnormal behavior can be detected by the OD mechanism of CoSec-RPL.  
\\
\textit{Wormhole attack detection:} To detect wormhole attacks, CoSec-RPL needs few modifications. First, the neighbor table needs to store the signal strength of its neighbors. Then, the neighbor with significantly strong signal strength can be classified as an attacker by CoSec-RPL. 
\\
\textit{\textcolor{black}{Spoofed copycat attack detection:}} \textcolor{black}{CoSec-RPL is designed specifically for detecting non-spoofed copycat attack. The present solution is not capable of detecting a spoofed copycat attack where an attacker may spoof its IP frequently. However, the attack detection logic of CoSec-RPL can be improved to detect such attacks. This limitation is left as an improvement to CoSec-RPL and will be considered in our future work.}

\section{Conclusion and Future scope}\label{Section:Conclusion}

RPL is currently the most popular routing protocol for 6LoWPAN based IoT applications. The security of such applications against various cyber attacks is one of the biggest challenges in the current scenario. In this paper, we first presented and investigated a routing attack named a copycat attack. The copycat attack is a combination of flooding and replay attack, which makes it \textcolor{black}{severe} for RPL based 6LoWPANs. From the simulation experiments, we illustrated how a \textcolor{black}{non-spoofed} copycat attack (i.e., a variant of copycat attack) significantly degraded the network performance, particularly in terms of AE2ED and PDR. We further, proposed and evaluated a distributed IDS named CoSec-RPL to secure 6LoWPAN against such \textcolor{black}{attacks}. Our proposed IDS detected \textcolor{black}{non-spoofed} copycat attack and showed acceptable performance in terms of ADA and FRT. Also, we have shown that the CoSec-RPL can be implemented on resource constrained node like the Zolertia Z1 mote. \textcolor{black}{The major limitations of CoSec-RPL include: (1) it cannot detect spoofed copycat attack; (2) it requires minor changes in RPL implementation; (3) it requires to maintain a neighbor table to store neighbor information.} In the future, we plan to improve CoSec-RPL performance and perform testbed experiments. 

\section*{ACKNOWLEDGMENT}
This research was supported by the Ministry of Human Resource Development, Government of India.

\section*{Conflicts of Interest
}
On behalf of all authors, the corresponding author states that there is no conflict of interest.

\bibliographystyle{spbasic} 
%\bibliography{mybibfile}

\begin{thebibliography}{72}
\providecommand{\natexlab}[1]{#1}
\providecommand{\url}[1]{{#1}}
\providecommand{\urlprefix}{URL }
\expandafter\ifx\csname urlstyle\endcsname\relax
  \providecommand{\doi}[1]{DOI~\discretionary{}{}{}#1}\else
  \providecommand{\doi}{DOI~\discretionary{}{}{}\begingroup
  \urlstyle{rm}\Url}\fi
\providecommand{\eprint}[2][]{\url{#2}}

\bibitem[{155(2018)}]{155IANA}
 (2018) {Internet Control Message Protocol version 6 (ICMPv6) Parameters}.
  \url{https://www.iana.org/assignments/icmpv6-parameters/icmpv6-parameters.xhtml},
  [Online; accessed 19-April-2018]

\bibitem[{Adat and Gupta(2018)}]{adat2018security}
Adat V, Gupta B (2018) Security in internet of things: issues, challenges,
  taxonomy, and architecture. Telecommunication Systems 67(3):423--441

\bibitem[{Airehrour et~al.(2016)Airehrour, Gutierrez, and
  Ray}]{AIREHROUR2016198}
Airehrour D, Gutierrez J, Ray SK (2016) Secure routing for internet of things:
  A survey. Journal of Network and Computer Applications 66:198 -- 213

\bibitem[{Airehrour et~al.(2018)Airehrour, Gutierrez, and Ray}]{Airehrour2018}
Airehrour D, Gutierrez JA, Ray SK (2018) {SecTrust -RPL: A secure trust-aware
  RPL routing protocol for Internet of Things}. Future Generation Computer
  Systems

\bibitem[{Alaba et~al.(2017)Alaba, Othman, Hashem, and Alotaibi}]{Alaba2017}
Alaba FA, Othman M, Hashem IAT, Alotaibi F (2017) {Internet of Things security:
  A survey}. Journal of Network and Computer Applications 88:10--28

\bibitem[{Ammar et~al.(2018)Ammar, Russello, and Crispo}]{ammar2018internet}
Ammar M, Russello G, Crispo B (2018) Internet of things: A survey on the
  security of iot frameworks. Journal of Information Security and Applications
  38:8--27

\bibitem[{Ancillotti et~al.(2018)Ancillotti, Bolettieri, and
  Bruno}]{ancillotti2018rtt}
Ancillotti E, Bolettieri S, Bruno R (2018) Rtt-based congestion control for the
  internet of things. In: International Conference on Wired/Wireless Internet
  Communication, Springer, pp 3--15

\bibitem[{Ashton(2009)}]{ashton2009internet}
Ashton K (2009) That `internet of things' thing. RFID journal 22(7):97--114

\bibitem[{Barnett and Lewis(1974)}]{barnett1974outliers}
Barnett V, Lewis T (1974) Outliers in statistical data. Wiley

\bibitem[{Bostani and Sheikhan(2017)}]{Bostani2017}
Bostani H, Sheikhan M (2017) {Hybrid of anomaly-based and specification-based
  IDS for Internet of Things using unsupervised OPF based on MapReduce
  approach}. Computer Communications

\bibitem[{Domingues et~al.(2018)Domingues, Filippone, Michiardi, and
  Zouaoui}]{domingues2018comparative}
Domingues R, Filippone M, Michiardi P, Zouaoui J (2018) {A comparative
  evaluation of outlier detection algorithms: Experiments and analyses}.
  Pattern Recognition 74:406--421

\bibitem[{Dunkels et~al.(2011)Dunkels, Schmidt, Finne, Eriksson, {\"O}sterlind,
  and Durvy}]{dunkels2011contiki}
Dunkels A, Schmidt O, Finne N, Eriksson J, {\"O}sterlind F, Durvy NTM (2011)
  The contiki os: The operating system for the internet of things.
  \url{http://www. contikios. org}, [Online; accessed 10-March-2020]

\bibitem[{Gaddour and Koubâa(2012)}]{RPLNutshell}
Gaddour O, Koubâa A (2012) {RPL in a nutshell: A survey}. Computer Networks
  56(14):3163 -- 3178

\bibitem[{Gara et~al.(2017)Gara, Saad, and Ayed}]{GaraInderscience}
Gara F, Saad LB, Ayed RB (2017) An efficient intrusion detection system for
  selective forwarding and clone attackers in ipv6-based wireless sensor
  networks under mobility. International Journal on Semantic Web and
  Information Systems (IJSWIS) 13(3):22--47

\bibitem[{Ghaleb et~al.(2018{\natexlab{a}})Ghaleb, Al-Dubai, Ekonomou, Qasem,
  Romdhani, and Mackenzie}]{ghaleb2018addressing}
Ghaleb B, Al-Dubai A, Ekonomou E, Qasem M, Romdhani I, Mackenzie L
  (2018{\natexlab{a}}) {Addressing the DAO Insider Attack in RPL’s Internet
  of Things Networks}. IEEE Communications Letters 23(1):68--71

\bibitem[{Ghaleb et~al.(2018{\natexlab{b}})Ghaleb, Al-Dubai, Ekonomou,
  Alsarhan, Nasser, Mackenzie, and Boukerche}]{ghaleb2018survey}
Ghaleb B, Al-Dubai AY, Ekonomou E, Alsarhan A, Nasser Y, Mackenzie LM,
  Boukerche A (2018{\natexlab{b}}) A survey of limitations and enhancements of
  the ipv6 routing protocol for low-power and lossy networks: A focus on core
  operations. IEEE Communications Surveys \& Tutorials 21(2):1607--1635

\bibitem[{Gnawali and Levis(2010)}]{ETXOF}
Gnawali O, Levis P (2010) {The ETX Objective Function for RPL,”
  draft-gnawali-roll-etxof-01}.
  \urlprefix\url{https://tools.ietf.org/html/draft-gnawali-roll-etxof-00}

\bibitem[{Gnawali and Levis(2012)}]{MRHOF}
Gnawali O, Levis P (2012) {The Minimum Rank with Hysteresis Objective
  Function}. Tech. rep., \urlprefix\url{https://tools.ietf.org/html/rfc6719}

\bibitem[{Gungor and Hancke(2009)}]{gungor2009industrial}
Gungor VC, Hancke GP (2009) Industrial wireless sensor networks: Challenges,
  design principles, and technical approaches. IEEE Transactions on industrial
  electronics 56(10):4258--4265

\bibitem[{Hamid et~al.(2006)Hamid, Rashid, and Hong}]{hamid2006routing}
Hamid MA, Rashid M, Hong CS (2006) Routing security in sensor network: Hello
  flood attack and defense. IEEE, Proceedings of First International Conference
  on Next-Generation Wireless Systems (ICNEWS), pp 2--4

\bibitem[{Hoaglin(2003)}]{hoaglin2003john}
Hoaglin DC (2003) {John W. Tukey and data analysis}. Statistical Science pp
  311--318

\bibitem[{Hui(2012)}]{hui2012routing}
Hui JW (2012) {The Routing Protocol for Low-Power and Lossy Networks (RPL)
  Option for Carrying RPL Information in Data-Plane Datagrams}
  \urlprefix\url{https://tools.ietf.org/html/rfc6553}

\bibitem[{IDC(2019)}]{IDC}
IDC (2019) {IDC Forecasts Worldwide Spending on the Internet of Things to Reach
  \$745 Billion in 2019, Led by the Manufacturing, Consumer, Transportation,
  and Utilities Sectors}.
  \url{https://www.idc.com/getdoc.jsp?containerId=prUS44596319}, [Online:
  accessed: 25-August-2019]

\bibitem[{IDC(2020)}]{idc2}
IDC (2020) {The Growth in Connected IoT Devices Is Expected to Generate 79.4ZB
  of Data in 2025, According to a New IDC Forecast}.
  \url{https://www.idc.com/getdoc.jsp?containerId=prUS45213219}, [Online;
  accessed 8-March-2020]

\bibitem[{Jabez and Muthukumar(2015)}]{jabez2015intrusion}
Jabez J, Muthukumar B (2015) {Intrusion detection system (IDS): anomaly
  detection using outlier detection approach}. Procedia Computer Science
  48:338--346

\bibitem[{Kabilan et~al.(2018)Kabilan, Bhalaji, Selvaraj, Kumaar, and
  Karthikeyan}]{kabilan2018performance}
Kabilan K, Bhalaji N, Selvaraj C, Kumaar M, Karthikeyan P (2018) {Performance
  analysis of IoT protocol under different mobility models}. Computers \&
  Electrical Engineering 72:154--168

\bibitem[{Kanaris et~al.(2019)Kanaris, Sergiou, Kokkinis, Pafitis, Antoniou,
  and Stavrou}]{kanaris2019realistic}
Kanaris L, Sergiou C, Kokkinis A, Pafitis A, Antoniou N, Stavrou S (2019) On
  the realistic radio and network planning of iot sensor networks. Sensors
  19(15):3264

\bibitem[{Kumar and Kumar(2016)}]{kumar2016anomaly}
Kumar N, Kumar U (2016) {Anomaly-Based Network Intrusion Detection: An Outlier
  Detection Techniques}. In: International Conference on Soft Computing and
  Pattern Recognition, Springer, pp 262--269

\bibitem[{Le et~al.(2013)Le, Loo, Lasebae, Vinel, Chen, and
  Chai}]{le2013impact}
Le A, Loo J, Lasebae A, Vinel A, Chen Y, Chai M (2013) The impact of rank
  attack on network topology of routing protocol for low-power and lossy
  networks. IEEE Sensors Journal 13(10):3685--3692

\bibitem[{Le et~al.(2016)Le, Loo, Chai, and Aiash}]{Le2016}
Le A, Loo J, Chai K, Aiash M (2016) {A specification-based IDS for detecting
  attacks on RPL-based network topology}. Information 7(2):25

\bibitem[{Levis et~al.(2011)Levis, Clausen, Hui, Gnawali, and
  Ko}]{levis2011trickle}
Levis P, Clausen T, Hui J, Gnawali O, Ko J (2011) The trickle algorithm. Tech.
  rep., \urlprefix\url{https://tools.ietf.org/html/rfc6206}

\bibitem[{Malik et~al.(2019)Malik, Dutta, and Granjal}]{malik2019survey}
Malik M, Dutta M, Granjal J (2019) {A Survey of Key Bootstrapping Protocols
  Based on Public Key Cryptography in the Internet of Things}. IEEE Access
  7:27443--27464

\bibitem[{Mayzaud et~al.(2014)Mayzaud, Sehgal, Badonnel, Chrisment, and
  Sch{\"{o}}nw{\"{a}}lder}]{Mayzaud2014}
Mayzaud A, Sehgal A, Badonnel R, Chrisment I, Sch{\"{o}}nw{\"{a}}lder J (2014)
  {A study of RPL DODAG version attacks}. Lecture Notes in Computer Science
  (including subseries Lecture Notes in Artificial Intelligence and Lecture
  Notes in Bioinformatics) 8508 LNCS:92--104

\bibitem[{Mayzaud et~al.(2015)Mayzaud, Sehgal, Badonnel, Chrisment, and
  Sch{\"o}nw{\"a}lder}]{mayzaud2015mitigation}
Mayzaud A, Sehgal A, Badonnel R, Chrisment I, Sch{\"o}nw{\"a}lder J (2015)
  {Mitigation of topological inconsistency attacks in RPL-based low-power lossy
  networks}. International Journal of Network Management 25(5):320--339

\bibitem[{Mayzaud et~al.(2016{\natexlab{a}})Mayzaud, Badonnel, and
  Chrisment}]{Mayzaud2016Version}
Mayzaud A, Badonnel R, Chrisment I (2016{\natexlab{a}}) {Detecting Version
  Number Attacks Using a Distributed Monitoring Architecture}. Proc of
  IEEE/IFIP/In Assoc with ACM SIGCOMM International Conference on Network and
  Service Management (CNSM 2016) pp 127--135

\bibitem[{Mayzaud et~al.(2016{\natexlab{b}})Mayzaud, Sehgal, Badonnel,
  Chrisment, and Sch{\"{o}}nw{\"{a}}lder}]{Mayzaud2016}
Mayzaud A, Sehgal A, Badonnel R, Chrisment I, Sch{\"{o}}nw{\"{a}}lder J
  (2016{\natexlab{b}}) {Using the RPL protocol for supporting passive
  monitoring in the Internet of Things}. Proceedings of the NOMS 2016 - 2016
  IEEE/IFIP Network Operations and Management Symposium pp 366--374

\bibitem[{Mayzaud et~al.(2017)Mayzaud, Badonnel, and Chrisment}]{Mayzaud2017}
Mayzaud A, Badonnel R, Chrisment I (2017) {A distributed monitoring strategy
  for detecting version number attacks in RPL-based networks}. IEEE
  Transactions on Network and Service Management 14(2):472--486

\bibitem[{Medjek et~al.(2018)Medjek, Tandjaoui, Romdhani, and
  Djedjig}]{medjek2018security}
Medjek F, Tandjaoui D, Romdhani I, Djedjig N (2018) Security threats in the
  internet of things: Rpl's attacks and countermeasures. In: Security and
  Privacy in Smart Sensor Networks, IGI Global, pp 147--178

\bibitem[{Mohammadi and Ghaffari(2019)}]{mohammadi2019defending}
Mohammadi P, Ghaffari A (2019) Defending against flooding attacks in mobile
  ad-hoc networks based on statistical analysis. Wireless Personal
  Communications 106(2):365--376

\bibitem[{Musaddiq et~al.(2018)Musaddiq, Zikria, Hahm, Yu, Bashir, and
  Kim}]{Musaddiq}
Musaddiq A, Zikria YB, Hahm O, Yu H, Bashir AK, Kim SW (2018) {A Survey on
  Resource Management in IoT Operating Systems}. IEEE Access 6:8459--8482

\bibitem[{Čolaković and Hadžialić(2018)}]{COLAKOVIC2018}
Čolaković A, Hadžialić M (2018) {Internet of Things (IoT): A review of
  enabling technologies, challenges, and open research issues}. Computer
  Networks 144:17 -- 39

\bibitem[{Pathan et~al.(2006)Pathan, Lee, and Hong}]{pathan2006security}
Pathan ASK, Lee HW, Hong CS (2006) Security in wireless sensor networks: issues
  and challenges. IEEE, 2006 8th International Conference Advanced
  Communication Technology, pp 1043--1048

\bibitem[{Perazzo et~al.(2017{\natexlab{a}})Perazzo, Vallati, Anastasi, and
  Dini}]{Perazzo2017}
Perazzo P, Vallati C, Anastasi G, Dini G (2017{\natexlab{a}}) {DIO Suppression
  Attack Against Routing in the Internet of Things}. IEEE Communications
  Letters 21(11):2524--2527

\bibitem[{Perazzo et~al.(2017{\natexlab{b}})Perazzo, Vallati, Arena, Anastasi,
  and Dini}]{Perazzo2017a}
Perazzo P, Vallati C, Arena A, Anastasi G, Dini G (2017{\natexlab{b}}) {An
  implementation and evaluation of the security features of RPL}. In: Lecture
  Notes in Computer Science (including subseries Lecture Notes in Artificial
  Intelligence and Lecture Notes in Bioinformatics)

\bibitem[{{Raoof} et~al.(2019){Raoof}, {Matrawy}, and {Lung}}]{Raoof}
{Raoof} A, {Matrawy} A, {Lung} C (2019) {Routing Attacks and Mitigation Methods
  for RPL-Based Internet of Things}. IEEE Communications Surveys Tutorials
  21(2):1582--1606

\bibitem[{Raymond and Midkiff(2008)}]{raymond2008denial}
Raymond DR, Midkiff SF (2008) Denial-of-service in wireless sensor networks:
  Attacks and defenses. IEEE Pervasive Computing (1):74--81

\bibitem[{Raza(2013)}]{Raza1043301}
Raza S (2013) Lightweight security solutions for the internet of things. PhD
  thesis, , SICS, \urlprefix\url{http://soda.swedishict.se/5548/}

\bibitem[{Raza et~al.(2013)Raza, Wallgren, and Voigt}]{RAZA20132661}
Raza S, Wallgren L, Voigt T (2013) Svelte: Real-time intrusion detection in the
  internet of things. Ad Hoc Networks 11(8):2661 -- 2674

\bibitem[{Seeber et~al.(2013)Seeber, Sehgal, Stelte, Rodosek, and
  Schonwalder}]{seeber2013towards}
Seeber S, Sehgal A, Stelte B, Rodosek GD, Schonwalder J (2013) {Towards a trust
  computing architecture for RPL in Cyber Physical Systems}. In: 2013 9th
  International Conference on Network and Service Management (CNSM), IEEE, pp
  134--137

\bibitem[{Sfar et~al.(2018)Sfar, Natalizio, Challal, and
  Chtourou}]{RIAHISFAR2018118}
Sfar AR, Natalizio E, Challal Y, Chtourou Z (2018) {A roadmap for security
  challenges in the Internet of Things}. Digital Communications and Networks
  4(2):118 -- 137

\bibitem[{Shamsoshoara et~al.(2019)Shamsoshoara, Korenda, Afghah, and
  Zeadally}]{shamsoshoara2019survey}
Shamsoshoara A, Korenda A, Afghah F, Zeadally S (2019) A survey on
  hardware-based security mechanisms for internet of things. arXiv preprint
  arXiv:190712525

\bibitem[{Thubert(2012)}]{OF0}
Thubert P (2012) Objective function zero for the routing protocol for low-power
  and lossy networks (rpl). Tech. rep.,
  \urlprefix\url{https://www.rfc-editor.org/info/rfc6552}

\bibitem[{Thulasiraman and Wang(2019)}]{thulasiraman2019lightweight}
Thulasiraman P, Wang Y (2019) {A Lightweight Trust-Based Security Architecture
  for RPL in Mobile IoT Networks}. In: 2019 16th IEEE Annual Consumer
  Communications \& Networking Conference (CCNC), IEEE, pp 1--6

\bibitem[{Tripathi(2014)}]{tripathi2014design}
Tripathi J (2014) On design, evaluation and enhancement of ip-based routing
  solutions for low power and lossy networks. PhD thesis, Drexel University

\bibitem[{Tsao et~al.(2015)Tsao, Alexander, Dohler, Daza, Lozano, and
  Richardson}]{tsao2015security}
Tsao T, Alexander R, Dohler M, Daza V, Lozano A, Richardson M (2015) A security
  threat analysis for the routing protocol for low-power and lossy networks
  (rpls). Tech. rep.,
  \urlprefix\url{http://dx.doi.org/10.1007/978-0-387-78875-3_9}

\bibitem[{Vallati(2019)}]{vallatiemail}
Vallati C (2019) personal communication

\bibitem[{Vasseur et~al.(2011)Vasseur, Agarwal, Hui, Shelby, Bertrand, and
  Chauvenet}]{IPSOAllianceRPL}
Vasseur J, Agarwal N, Hui J, Shelby Z, Bertrand P, Chauvenet C (2011) {RPL: The
  IP routing protocol designed for low power and lossy networks}. Internet
  Protocol for Smart Objects (IPSO) Alliance 36,
  \urlprefix\url{http://www.ipso-alliance.org/wp-content/media/rpl.pdf}

\bibitem[{Verma and Ranga(2019{\natexlab{a}})}]{verma2019addressing}
Verma A, Ranga V (2019{\natexlab{a}}) {Addressing Flooding Attacks in
  IPv6-based Low Power and Lossy Networks}. In: TENCON 2019-2019 IEEE Region 10
  Conference (TENCON), IEEE, pp 552--557

\bibitem[{Verma and Ranga(2019{\natexlab{b}})}]{abhishek_verma_IoTSIU}
Verma A, Ranga V (2019{\natexlab{b}}) {ELNIDS: Ensemble Learning based Network
  Intrusion Detection System for RPL based Internet of Things}. In: 2019 4th
  International Conference on Internet of Things: Smart Innovation and Usages
  (IoT-SIU), IEEE, pp 1--6

\bibitem[{Verma and Ranga(2019{\natexlab{c}})}]{Verma2019}
Verma A, Ranga V (2019{\natexlab{c}}) {Evaluation of Network Intrusion
  Detection Systems for RPL Based 6LoWPAN Networks in IoT}. Wireless Personal
  Communications pp 1--24

\bibitem[{Verma and Ranga(2020)}]{verma2020mitigation}
Verma A, Ranga V (2020) {Mitigation of DIS flooding attacks in RPL-based
  6LoWPAN networks}. Transactions on Emerging Telecommunications Technologies
  31(2):e3802

\bibitem[{{Verma} and {Ranga}(2020)}]{VermaIEEE}
{Verma} A, {Ranga} V (2020) {Security of RPL based 6LoWPAN Networks in the
  Internet of Things: A Review}. IEEE Sensors Journal
  \doi{10.1109/JSEN.2020.2973677}, {Early Access}

\bibitem[{Wadhaj et~al.(2020)Wadhaj, Ghaleb, Thomson, Al-Dubai, and
  Buchanan}]{wadhaj2020mitigation}
Wadhaj I, Ghaleb B, Thomson C, Al-Dubai A, Buchanan WJ (2020) {Mitigation
  Mechanisms Against the DAO Attack on the Routing Protocol for Low Power and
  Lossy Networks (RPL)}. IEEE Access 8:43665--43675

\bibitem[{Wallgren et~al.(2013)Wallgren, Raza, and Voigt}]{wallgren2013routing}
Wallgren L, Raza S, Voigt T (2013) {Routing Attacks and Countermeasures in the
  RPL-based Internet of Things}. International Journal of Distributed Sensor
  Networks 9(8):794326

\bibitem[{Wang et~al.(2018)Wang, Li, Fang, and Wang}]{wang2018robust}
Wang H, Li H, Fang J, Wang H (2018) {Robust Gaussian Kalman filter with outlier
  detection}. IEEE Signal Processing Letters 25(8):1236--1240

\bibitem[{Wang et~al.(2017)Wang, Chalhoub, Tall, and Misson}]{wang2017routing}
Wang J, Chalhoub G, Tall H, Misson M (2017) Routing protocol enhancement for
  mobility support in wireless sensor networks. In: International Conference on
  Ad-Hoc Networks and Wireless, Springer, pp 262--275

\bibitem[{Winter et~al.(2012)Winter, Thubert, Brandt, Hui, Kelsey, Levis,
  Pister, Struik, Vasseur, and Alexander}]{winter2012rpl}
Winter T, Thubert P, Brandt A, Hui J, Kelsey R, Levis P, Pister K, Struik R,
  Vasseur JP, Alexander R (2012) {RPL: IPv6 routing protocol for low-power and
  lossy networks}. Tech. rep.,
  \urlprefix\url{https://tools.ietf.org/html/rfc6550}

\bibitem[{Xie et~al.(2010)Xie, Goyal, Hosseini, Martocci, Bashir, Baccelli, and
  Durresi}]{xie2010routing}
Xie W, Goyal M, Hosseini H, Martocci J, Bashir Y, Baccelli E, Durresi A (2010)
  Routing loops in dag-based low power and lossy networks. In: Advanced
  Information Networking and Applications (AINA), 2010 24th IEEE International
  Conference on, IEEE, pp 888--895

\bibitem[{Xu et~al.(2014)Xu, He, and Li}]{IoTCore1}
Xu LD, He W, Li S (2014) {Internet of Things in Industries: A Survey}. IEEE
  Transactions on Industrial Informatics 10(4):2233--2243

\bibitem[{Yang et~al.(2017)Yang, Wu, Yin, Li, and Zhao}]{Yang7902207}
Yang Y, Wu L, Yin G, Li L, Zhao H (2017) {A Survey on Security and Privacy
  Issues in Internet-of-Things}. IEEE Internet of Things Journal
  4(5):1250--1258

\bibitem[{Zhi et~al.(2018)Zhi, Luo, and Liu}]{zhi2018gini}
Zhi T, Luo H, Liu Y (2018) {A Gini impurity-based interest flooding attack
  defence mechanism in NDN}. IEEE Communications Letters 22(3):538--541

\bibitem[{Ziegeldorf et~al.(2014)Ziegeldorf, Morchon, and
  Wehrle}]{ziegeldorf2014privacy}
Ziegeldorf JH, Morchon OG, Wehrle K (2014) Privacy in the internet of things:
  threats and challenges. Security and Communication Networks 7(12):2728--2742

\end{thebibliography}

\end{document}